\documentclass[review]{elsarticle}

\usepackage{mathtools}
\usepackage{graphicx}
\usepackage{xcolor}
\usepackage{multicol}
\journal{Journal of \LaTeX\ Templates}

\begin{document}

\begin{frontmatter}

\title{An Efficient Mechanism for Computation Offloading in Mobile-Edge Computing}



\author[mymainaddress]{Mahla Rahati-Quchani \corref{mycorrespondingauthor}}
\cortext[mycorrespondingauthor]{Corresponding author}
\ead{mahla.rahati@mail.um.ac.ir }

\author[mymainaddress]{Saeid Abrishami \corref{mycorrespondingauthor}}
\ead{s-abrishami@um.ac.ir}

\author[mysecondaryaddress]{Mehdi Feizi \corref{mycorrespondingauthor}}
\ead{feizi@um.ac.ir}

\address[mymainaddress]{Department of Computer Engineering, Engineering Faculty, Ferdowsi University of Mashhad, Azadi Square, Mashhad, Iran}
\address[mysecondaryaddress]{Department of Economics, Ferdowsi University of Mashhad, Azadi Square, Mashhad, Iran}

\begin{abstract}
Mobile edge computing (MEC) is a promising technology that provides cloud and IT services within the proximity of the mobile user. With the increasing number of mobile applications, mobile devices (MD) encounter limitations of their resources, such as battery life and computation capacity. The computation offloading in MEC can help mobile users to reduce battery usage and speed up task execution. Although there are many solutions for offloading in MEC, most usually only employ one MEC server for improving mobile device energy consumption and execution time. Instead of conventional centralized optimization methods, the current paper considers a decentralized optimization mechanism between MEC servers and users. In particular, an assignment mechanism called school choice is employed to assist heterogeneous users to select different MEC operators in a distributed environment. With this mechanism, each user can benefit from minimizing the price and energy consumption of executing tasks while also meeting the specified deadline. The present research has designed an efficient mechanism for a computation offloading scheme that achieves minimal price and energy consumption under latency constraints. Numerical results demonstrate that the proposed algorithm can attain efficient and successful computation offloading.
\end{abstract}


\end{frontmatter}


\section{Introduction}

Nowadays, mobile devices face some restrictions due to the rapid development of mobile applications, such as those for face recognition, natural language processing, interactive gaming, and augmented reality. These applications are usually latency sensitive, computationally intensive, and high in the energy consumption, such that MDs with limited battery and computational resources can hardly support such programs. The European Telecommunications Standards Institute (ETSI) has introduced mobile edge computing as a new standardization group. The purpose is to provide information technology and cloud computing capabilities in the proximity of mobile users. In this way, users are offered a service environment characterized by proximity, low latency, high rate access, and sufficient resources that combine edge computing, mobile devices, and a wireless network. As a result, users can take advantage of more intelligent applications\cite{Hu2015b}. 

The field of computation offloading research addresses the sending of tasks to MEC servers, which can be a small data center in the users’ area created by the telecom operator.  More exactly, with less than one-millisecond standard latency, MEC and 5G facilitate the usage of cloud resources in the proximity of MDs and so can effectively support delay sensitive applications. In comparison to MDs, the mobile edge offers many significant advantages, including servers providing more computational resources which enable applications to run faster and more efficiently. Connected with one hop, MEC servers consume less energy and time than a cloud in the sending and receiving of application data. In a distributed geographic area, disparate operators have several servers with different and limited computational resources. These may serve many MDs with endless sequences of computational tasks, various application characteristics, and varied communication requirements. Therefore, multiple heterogeneous MDs compete for numerous heterogeneous MEC resources\cite{Li2018Users}.

Computation offloading in MEC is an essential technique to allow users to access computational capabilities at the network edge. Each user can decide to offload a computational task instead of running it locally. Since the primary purpose of offloading is to reduce the energy consumption and execution time, most previous works have considered these two parameters \cite{Zhang2016,  Zhao2017, Zhang2017a, Hao2018, Zhang2018, You2017, Li2019a, 8241344, Guo2018, Dai2018a,  Tran2017, Dinh2017, Chen2018b, Ugwuanyi2018, Huang, Li2019c}. However, for MEC operators, the usage of resources is costly. As a result, more recent works have focused on the monetary revenue of operators \cite{Nguyen2018, Gao2019, 8166725,  Ranadheera2018, Li2018f}. Few researchers have considered all of these parameters at the same time\cite{Zhang2017d, Gu2018, Gu2018a}.

Numerous studies have solved the computation offloading problem from the standpoint of one single service operator, thus indicating that the authors employed only one MEC with a centralized manager \cite{Zhao2017, Zhang2016, You2017, Ren2018}. However, the most recent works have investigated the possibility of multiple MECs within the user area. With its significant overhead and complexity, a centralized controller for multi-MEC environments is less applicable. As a result, most recent works have concentrated on decentralized methods \cite{Yang2018, Dinh2018, Bahreini2018, Sun2018}.

Game theory has been one of the most powerful tools employed to tackle this problem \cite{Chen2016, Guo2018b, Yang2018cooperative, Dinh2018, Li2018f, Li2018, Li2018Users, Yi2019, Zhang2017, 8249785, Yang2018}. This is a useful framework to analyze interactions among independent MDs acting in their interests. Users can make offloading decisions for their benefit and play the game until reaching a stable state (i.e., Nash Equilibrium). Without a central authority, such decisions can be reached based on local information about the system and environment conditions, thus preventing information collection from massive mobile devices by a central controller.

In mechanism design, a market maker generally assumes the preference of supposedly rational agents seeking to promote their own benefit and matches each of them with another agent/object. However, since the market maker is interested in certain characteristics of the final match and depends on these, he or she might choose different matching mechanisms. Most previous works on MEC mechanisms have concentrated on the stability of the final matching as its desired characteristic despite having sacrificed efficiency  \cite{Gu2018, Pham2018a, Gu2018a}. Even so, efficiency plays a critical role in MEC because of the necessity for quick responses.  In our specific case, we are looking for a matching mechanism which guarantees efficiency.

Furthermore, the matching theory considers both sides, i.e. MECs and MDs, as agents with preferences. In this paper, although MDs may prefer cheaper and lower latency MECs, but the MECs prioritize users based on distance. The present study holds that MECs have no preferences on MDs, even though MECs have a higher priority for MD's nearer to them. Consequently, MECs can be considered as objects and the current work may utilize the concept of assignment mechanism instead of matching. In the assignment mechanism, there are two sets of agents and objects which should be assigned to each other. The difference between these two sets is that agents have preferences over objects, while objects only have priorities on agents \cite{Haeringera}. MECs do not have preferences over MDs, but they nevertheless rank them. To differentiate these rankings over MDs from the concept of preferences, will call such rankings priorities or priority lists. In real life, those priorities are often the outcome of technical constraints. For instance, MDs who are located close to a MEC have a higher priority for connecting to that MEC than the MDs who are farther away. One may think that the MECs "prefers" MDs who are closer, although it is better to say that the MDs who are closer have a higher priority than the MDs who are not.

To the best of our knowledge, this work is the first attempt to employs one of the well-known assignment mechanisms, the school choice, to tackle the problem of computation offloading in a multi-user and multi-MEC environment. An efficient but unstable algorithm is used to meet the low latency requirements. The proposed algorithm aims to minimize the price and energy consumption of the user task while meeting its deadline. In comparison with energy based offloading, price based offloading, and the heuristic offloading decision algorithm (HODA) proposed in \cite{Lyu2017a}, simulation results show that the current study’s algorithm outperforms these in terms of minimize the price and energy consumption of the user task and increas the percentage of successful offloading users.

The rest of the paper is organized as follows. Section II presents related works on computation offloading technologies in MEC. Section III introduces the system model of this scheme. Section IV describes the school choice for the computation offloading algorithm. Section V discusses the simulation results and, finally, Section VI concludes the work.

\section{Related Works}

The objective of the current research is to study the MEC offloading problem. A set of studies have already focused on how to optimize MEC task offloading from different perspectives. In fact, there are plenty of works in the area of computation offloading in MEC environments. In terms of architecture, most early studies consider single-user and single-MEC environments \cite{Zhang2018, Hao2018, Zhao2017, Zhang2017a, Zhang2016}. However, most recent works study the multi-user and multi-MEC \cite{Ugwuanyi2018, Yang2018, Dinh2018}. In terms of method, these studies are divided into two categories, namely centralized offloading and decentralized offloading approaches.

\subsection{Centralized Offloading Approaches}

In the centralized method, all information must be sent to a controller which makes the decisions. From the start, most works have focused on single-user and simple scenarios \cite{Mao2016e, Dinh2017, Wang2016a}, but these approaches are not practical in a real world with multiple users.

Zhang et al. explored a new multi-user scenario with one MEC. To minimize energy consumption, they formulated an optimization problem in which the energy cost of both task computing and file transmission are taken into consideration \cite{Zhang2016}. Zhao et al. designed other multi-users and single-MECs, which jointly optimize the offloading selection, radio resource allocation, and computational resource allocation \cite{Zhao2017}.

Because of the limited MEC servers, many users cannot offload in each time slot. So, Tang et al. addressed a scenario about maximizing the offloaded tasks number in MEC. They analyzed and solved the partial offloaded task number maximization problem by using the block coordinate descent (BCD) method. Also, they investigated their method to UAV (Unmanned Aerial Vehicle) enabled MEC system. The solution idea of UAV enabled MEC system’s optimization problem is the same as first method. Therefore, their partial offloading strategy can be seamlessly adapted to UAV enabled MEC system \cite{Tang2020}.

You et al. in \cite{You2017} proposed a scenario in which mobile users have different computation workloads and local computation capacities. In addition, they formulated a convex optimization problem with partial offloading to minimize the sum of mobile-energy consumption. Their key finding was that the optimal policy for controlling offloading data size and time allocation has a simple threshold-based structure. Besides, this study assumed that the edge has perfect knowledge of the local computing energy consumption, channel gains, and fairness factors of all mobile users. This information is utilized for the design of a centralized resource allocation to achieve the minimum weighted sum of mobile energy consumption. This result was also extended to OFDMA-based MEC systems, which offer orthogonal frequency-division multiple access for devising a near-optimal computation offloading policy.

To minimize the latency of all devices under limited communication and computation resource constraints, Ren et al. designed a multi-user video compression offloading. They studied and compared three models: local compression, edge cloud compression, and partial compression offloading. For the local compression model, a convex optimization problem was formulated to minimize the weighted-sum delay of all devices under the communication resource constraint. They considered that massive online monitoring data should be transmitted and analyzed by a central unit. For the edge cloud compression model, this work analyzed the task completion process by modeling a joint resource allocation problem with the constraints of both communication and computation resources. For the partial compression offloading model, they first devised a piecewise optimization problem and then derived an optimal data segmentation strategy in a piecewise structure. Finally, numerical results demonstrated that the partial compression offloading can more efficiently reduce end-to-end latency in comparison with the two other models \cite{Ren2018}.

A centralized computational offloading model may be challenging to run when massive offloading information is received in real time. As errors during the data gathering step may produce inefficient results, the local mode is more reliable and accurate than centralized solutions. Therefore, in many cases, the results of distributed approaches are more robust than those of centralized solutions. Due to the computational complexity of the scenario and numerous data from independent MDs, computation offloading for multi-user and multi-MEC systems poses a great challenge in a centralized environment \cite{Han2017a}.

\subsection{Decentralized Offloading Approach}

With the development of MEC, data traffic has rapidly grown in recent years. In response to this, offloading for MDs has been considered as an optimal solution. Traditional centralized offloading approaches cannot meet the requirements of emerging interactive programs for long-term communications. Consequently, the distributed computation model is preferred. Especially in a multi-server scenario, information gathering and decision-making do not require a central controller and each user can choose to offload as desired. Furthermore, distributed computation offloading models are flexible and scalable.

\subsubsection{Heuristic methods}

Some practical studies have investigated the task offloading problem in order to minimize execution time and energy consumption. For the distributed computation offloading model, Ugwuanyi et al. \cite{Ugwuanyi2018} proposed a resource provisioning algorithm based on the Banker’s algorithm. This was a way to provide higher reliability of network interactions that utilize software-defined networking for the reduction of communication overhead. Because edge nodes have a finite amount of resources, this study attempted to avoid a deadlock situation. Furthermore, since a deadlock may occur due to the considerable number of devices contending for a limited amount of resources, they also considered overdemand and delays while provisioning resources.

Tran et al. proposed an algorithm for a multi-MEC and multi-user scenario which minimizes the latency and energy consumption by optimizing the task offloading decision, uplink transmission power, and computational resource allocation. However, their algorithm does not guarantee the latency requirements for all users \cite{Tran2017}.

The heuristic offloading decision algorithm (HODA) proposed in \cite{Lyu2017a} is semi-distributed and runs in two stages. In the first stage, each mobile user independently optimizes the transmission power and determines whether to send an offloading request. In the second stage, the macro cell locally forms an optimal offloading set by prioritizing users according to the maximum utility. Finally, the selected mobile users offload their computation tasks. In this paper, the resource providers can either increase the capability of the MEC computing center or deploy multiple computing centers. In the case of multiple computing centers, the mobile users can apply the offloading policies that make offloading decisions among multiple providers independently and send offloading requests to the selected computing center.

\subsubsection{Game theory}

Recent advances in technology and the ever-increasing demand for computing and communications have generated an urgent need for a new analytical framework to address the present and future technical challenges of wireless and communication networks. As a result, game theory has been recognized in recent years as a central tool for the design of wireless networks and future communications. The Game theory arose out of the necessity to combine the rules and techniques of decision-making for future generations of wireless and communication nodes. Thus, depending on their requirements for various services, users can effectively communicate, for example, by video streaming over mobile networks. In brief, game theory offers excellent benefits for future wireless networks, such as decision making based on local information and distributed implementation, robust results, appropriate approaches for solving problems of a combinatorial nature, and rich mathematical and analytical tools for optimization \cite{Han2012}. Lately, game theory has been widely employed as a powerful tool for the distributed computation offloading approach among multiple MDs with varied interests.

In this context, Yi et al. scheduled joint computation offloading and wireless transmission scheduling with delay-sensitive applications in a single MEC scenario. In this model, there is a base station with multi-channels and multi-users who send a request to each channel. If there is more than one request for one channel, a queuing model is required for dynamic management of the computation offloading and transmission scheduling for mobile edge computing. This study formulated a novel mechanism, named MOTM, to jointly decide computation offloading, transmission scheduling, and a pricing rule \cite{Yi2019}. Nevertheless, single-MEC is a simple scenario and there are currently several servers within a user area from which to choose.

With a multi-user and multi-MEC scenario in mind, Yang et al. designed a potential game between a multi-user and multi-MEC for a distributed computation offloading approach. They solved the total overhead in terms of latency and energy consumption problem which consist of the weighting parameters affecting total system overhead: latency, interference, and energy consumption\cite{Yang2018}.

For multi-server and multi-user networks, Dinh et al. in \cite{Dinh2018} proposed a reinforcement learning offloading mechanism (Q-learning) to achieve long-term utilities. They modeled a theoretical game framework in which MDs select their targeted edge nodes and the transmission power to maximize their processed CPU cycles in each time slot, while also saving on energy consumption. Unlike existing works requiring predefined stochastic dynamics of channels for learning strategies, this study adopted a model-free reinforcement learning mechanism to design an offloading policy for players. 

%

\subsubsection{Mechanism design}

Mechanism design has some distinctive features. For example, a game “designer” chooses the game structure rather than inheriting one. Therefore, the mechanism design is often called “reverse game theory." Also, the designer is interested in the game’s outcome \cite{Han2012}. In computational offloading problems, the mechanism design has two important fields, auctions, and matching theory.

Some works only focus on the user's side to reduce the running time and energy consumption of the user device for offloading. However, other studies concentrate on the financial aspects of the operator's side since the use of resources incurs a cost for operators in terms of power consumption or other expenditures. Some authors, such as \cite{Bahreini2018, Sun2018, Zhang2017d}, have solved the issue by an auction. 

Bahreini et al. in \cite{Bahreini2018} proposed an envy-free auction mechanism for resource allocation in the MEC and cloud, in which users place bids for using a certain amount of resources. In this work, welfare was maximized when servers with different capacities in the edge or cloud and heterogeneous users competed for resources. The authors demonstrated that the proposed mechanism was individually-rational and returned envy-free allocations. Sun et al. also formulated their problem as an auction-based mechanism to address the computing resource allocation issue. They explored two double auction schemes with dynamic pricing in the MEC: a breakeven-based double auction (BDA) and a more efficient dynamic pricing-based double auction (DPDA). Under locality constraints, the schemes determined the matched pairs between MDs and edge servers, as well as the pricing mechanisms for high system efficiency \cite{Sun2018}. 

Considering heterogeneous requests, Zhang et al. in \cite{Zhang2017d} deployed the matching problem between MEC service providers (SPs) and MDs, which is a combinatorial auction-based service provider selection with limited wireless and computational resources. They modeled the matching relationship between MECs and MDs as a commodity trading process, which was offered a multi-round sealed sequential combinatorial auction (MSSCA) mechanism to match the SPs to MDs. In the one-round auction, the bidder cannot decide from which seller to buy and there must also be a central controller or coordination among sellers. This was not practical for this study’s network of multiple SPs and MDs without a centralized controller. As a result, the multi-round auction was applied to their auction design. 

Some other works have utilized a matching mechanism to solve computational offloading problems \cite{Gu2018, Pham2018a, Gu2018a}. In their paper \cite{Gu2018}, Gu et al. proposed a matching model called the Student Project Allocation (SPA), by which various students are assigned different projects belonging to different lecturers. In fog computing, they modeled the resource allocation problem as the SPA game, in which lecturers propose projects and students request these projects. Similarly, SPs offered available radio and CPU resource bundles, while users requested acceptable resource bundles from the SPs. SPs based their decisions on the revenue that could be generated from user requests for resource bundles.

In order to minimize the total computation overhead in terms of execution time and energy consumption, Pham et al. in \cite{Pham2018a} considered an optimization problem for jointly determining the computation offloading decision and allocating the offloading power at MEC servers. The computation offloading decision problem was divided into two subproblems for choosing between MEC servers and deciding on subchannels. To adopt these approaches, they employed the stable marriage problem which was two matching games with two sets of matching players. The first many-to-one matching game consisted of users with servers and the second one-to-one matching game had users with subchannel pairs. 

Recently, Gu et al. in \cite{Gu2018a} has also started to study offloading in the MEC with multi-user and multi-MEC and proposed a decentralized task offloading strategy between MDs and MEC servers. The task offloading problem was formulated as a one-to-many matching game to reduce overall energy consumption and terminated as a stable matching between tasks and nodes. These authors focused on the three parameters of energy consumption, execution time, and price. Also, they considered mobile devices with the excessive computational capability and MEC servers as edge nodes. Moreover, the total edge energy consumption for task execution was the sum of the transmission energy consumption and computation energy in the MEC.

In summary, for the multi-user and multi-MEC scenario, most existing works do not provide an efficient approach that reduces energy consumption and price under latency constraints. In contrast, the current work considers the distance of users and the benefits of the users when users and MEC operators want to select each other. In the scenario developed by the present study, once MEC servers have performed user tasks before the deadline and so have saved energy, they can earn money in exchange for providing resources to closer users. This scenario is efficient with multi-user and multi-MEC, which can select among operators with different prices in a decentralized environment to achieve minimal energy consumption and price under latency constraints.

\begin{figure}[t]
\centering
\includegraphics[width=9cm]{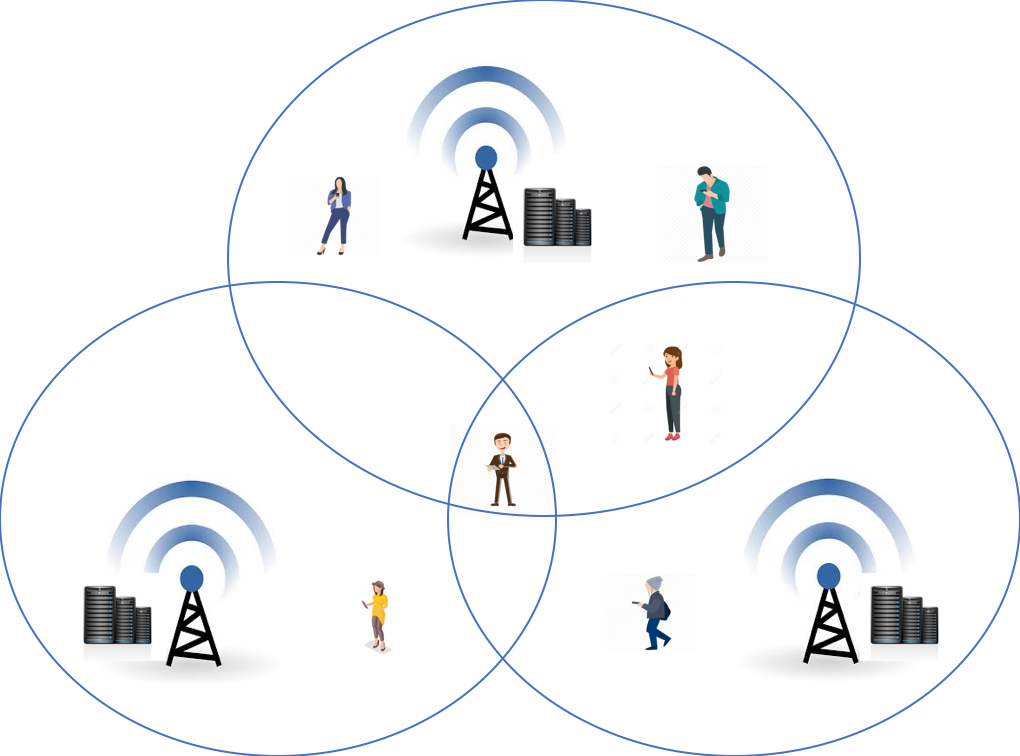}
\caption{System architecture}
\label{MEC.png}
\end{figure}

\section{System Model}

This section describes a system model adopted by the present study. As shown in Fig. \ref{MEC.png}, a network is assumed with multiple MDs denoted as $U= \{u_1, …, u_n, ..., u_N\}$ with a finite amount of money. Each MD $i$ has a different computational capability $f^{local}_i$ with a computationally intensive and a delay sensitive task featuring a different workload with a different deadline and users wanting to finish tasks before the deadline. The tasks include augmented reality, a health monitoring application, and an infotainment application, all of which differ in some properties, such as data upload or download. For the heterogeneous computing task, $t_i$, some properties are defined, such as $t_i$= ($d_i$, $b_i$, $T_i^{max}$), in which $d_i$ (in MI) is the amount of computing resource required for task $t_i$ and $b_i$ (in bits) is the data size of computation task $t_i$, that is, the amount of data content (for example, the processing code and parameter). Also, the task is not divisible and must be completed before deadline $T_i^{max}$. It should be noted that these properties are inherent parameters of the application and will not change according to where the application is processed.

In different areas, it is assumed that the edge computing system consists of multi-MEC donated as $S=\{s_1, ..., s_m, ..., s_M\}$. In the current work, each MEC server, $j$, has one host with different limited computational capabilities and is measured with million instructions per second (MIPS) shared among users. For each MEC server $j$, $f_j^{max}$ indicates the maximum computational capability shared among users and $q_{j}$ denotes the maximum number of users which the  MEC server can enroll. Many articles, such as \cite{Pham2018a}, have assumed that MEC servers have limited resources divided equally among users. However, equal allocation of resources is not appropriate because the user requirement may be more or less than the amount allocated.

In this scenario, many MEC servers are placed in the user location and users must decide which server is the best for offloading so as to achieve a successful process with a variant deadline. $\lambda_j$ is the unit price per CPU cycle to be paid by MD $i$ for the computation capability provided by MEC $j$. The present work assumes that everyone should pay the selected operator based on their consumption. Therefore, the user price is determined by $ \lambda_j{d_i} $. This price does not affect the selection by MEC servers because these servers initially serve closer users. Operators decide whether or not to select MD $i$ as their servers according to the server resource capacity $f^{max}_j$, and the distance to MD $i$.

Besides, the offloading decision for each computational task $t_i$ is denoted by $x_{ij} $, which takes zero and one. If MD $i$ is chosen to offload a task to MEC server $j$, then $x_{ij}$ must be equal to one or else $x_{ij}$ is zero, thus indicating that task $t_i$ is locally executed. A quasi-static scenario is considered when the set of users remains unchanged during a computation offloading period (e.g., several hundred milliseconds), even though this may change across different periods\cite{Chen2016}. Since both communication and computation play a key role in mobile edge computing, the next section discusses communication and computation models in detail.

\subsection{Communication Model}
Let $H_{ij}$ be the channel gain between MD $i$ and MEC server $j$ and show $P_i$ as the transmission power of MD $i$. According to the Shannon-Hartley formula, the achievable rate of MD $i$ can then be obtained as follows: 

\begin{equation}
\label{eq:uplink data rate}
r_i = B_{ij} \log_{2}(1 +  \frac{P_i H_{ij}}{\sigma ^ 2}  )
\end{equation}

where $\sigma ^ 2$ denotes the noise power and MD $i$ has bandwidth $B_{ij}$ representing the channel bandwidth of MEC $j$. The total bandwidth of $B_j$ should be equally divided among the MDs, such that each MD $i$ can obtain a non-overlapping frequency to simultaneously offload its data to the edge \cite{8166725},\cite{8249785},\cite{Ren2018}. In this paper, the downlink transmission delay is not considered because the size of the results is often smaller than the input data and the downlink rate from the MEC server $j$ to the MD $i$ is higher than the uplink rate from the MD $i$ to the MEC server $j$\cite{ Hao2018}, \cite{Mao2016e}.

\subsection{Computation Model}

The following discusses the computation overhead in terms of both energy consumption and the processing time for both local computing and edge computing approaches.

\subsubsection{Local Computing}

 For execution of local tasks, data need not be transmitted, but computation energy and processing time should be considered. According to the work of  \cite{Yang2018}, $\epsilon^{local}_i$ denotes the energy consumption coefficient per CPU cycle of MD $i$. Thus, the computational energy in local computing is:

\begin{equation}
\label{eq:LocalComputationalEnergy}
E^{local}_i =  \epsilon^{local}_i (d_i)
\end{equation}

The local processing time $T^{local}_i$ is computed as:

\begin{equation}
\label{eq:LocalExecTime}
T^{local}_i =  \frac{d_i}{f^{local}_i}
\end{equation}

\subsubsection{Edge Computing}
In the offloading mode, when the MEC remotely carries out the task, the MD is idle. However, an extra cost is necessary to send the input data for calculation. Hence, the transmission time $T^{trans}_i$ and transmission energy $E^{trans}_i$ for MD $i$ in the edge computation is: 

\begin{equation}
\label{eq:TransmissionTime}
\begin{aligned}
T^{trans}_i =\frac{b_i}{r_i}
\end{aligned}
\end{equation}

\begin{equation}
\label{eq:TransmissionEnergy}
\begin{aligned}
E^{trans}_i = \frac{b_i P_i}{r_i}
\end{aligned}
\end{equation}

The processing time of tasks in a MEC server is calculated by:

\begin{equation}
\label{eq:execTime}
\begin{aligned}
T^{exe}_i = \frac{d_i}{f_{ij}}
\end{aligned}
\end{equation}

In fact, $T_i = T^{exe}_i + T^{trans}_i$ in the edge computing mode and $T_i = T^{local}_i$ in the local computing mode. Based on the deadline, workload, and transmission time, each MD $i$ with distance $D_{ij}$ to MEC server $j$, determines the amount of resources it needs, which is denoted by $f_{ij}$. For this purpose, each task $t_i$ has a deadline with a different workload. Consequently, $f_{ij}$ can be calculated by considering the sum of transmission time $T^{trans}_i$, the deadline of offloading task $T^{max}_i$, and the workload of task $d_i$.  

\begin{equation}
\label{eq:required resource}
\begin{multlined}
f_{ij}= \frac{d_i}{T^{max}_i   -  T^{trans}_i }
\end{multlined}
\end{equation}

Based on this information, MD $i$ determines the required resource $f_{ij}$ on MEC server $j$. Since for each MD $i$, two parameters, namely the energy consumption and price received by the MEC, are involved in the selection of servers. So to be able to prioritize the servers, each user $i$ has a $\alpha_i$ and $\beta_i$ parameter that indicates the importance of the price and energy consumption for users.

\subsection{Problem Formulation}

The main goal in this problem is to minimize the energy consumption and price due to the task features and resources of the servers available to the user. Based on the user's decision, the total energy consumed by each task $t_i$ is equal to:

\begin{equation}
\label{eq:total energy}
\begin{multlined}
E_i = (x_{ij}  * E^{trans}_i) +  ((1 - x_{ij})* E^{local}_i)
\end{multlined}
\end{equation}

To minimize the energy consumption of the task execution and price, the utility function is :

\begin{equation}
\label{eq:utility function}
\begin{multlined}
U_i = (\alpha_{i}  * \lambda_j{d_i} ) + 
(\beta_{i}   *  E_i ) 
\end{multlined}
\end{equation}

the optimization problem can be formulated as follows: 

\begin{equation}
\label{eq:objective function}
\begin{multlined}
\hspace{1.2em}Min\hspace{1.2em} \displaystyle\sum_{i=1}^{N} U_i     \\
\hspace{-2em}\hspace{-1em}
s.t. \hspace{1em}
C1:T_i  \le T_i^{max}                                               \hspace{2em}          \forall i \in N                        \\
\hspace{3em}C2:    \displaystyle\sum_{i=1}^{N} x_{ij}      \le q_{j}                \hspace{1.5em}       \forall i \in N    ,\forall j \in M \\ 
\hspace{3.5em}C3:    f_{ij} > 0                                                       \hspace{3.8em}        \forall i \in N    ,\forall j \in M \\ 
\hspace{3.5em}C4:    x_{ij} = \{0,1\}                                                  \hspace{1.8em}        \forall i \in N    ,\forall j \in M \\    
\hspace{3.5em}C5:    \displaystyle\sum_{i=1}^{N} f_{ij}      \le f_j^{max}       \hspace{0.7em}        \forall i \in N      , \forall j \in M \\
\hspace{3.8em}C6:    \displaystyle\sum_{j=1}^{M} x_{ij}      \le 1                   \hspace{2.5em}        \forall i \in N      , \forall j \in M \\
\end{multlined}
\end{equation}

Here, the total MD energy consumption in the local execution or edge execution must be minimized. Completion of tasks is ensured before the deadline in the first constraint. Constraint C2 specifies that the total number of people offloading their tasks will not exceed the capacity of the MEC servers. Constraints C3 and C4 first assure that the computational capability allocated to each task $t_i$ is more than zero and that the MD's decisions have two modes: local execution and edge execution. Constraint C5 demonstrates that the total capacity received by the MDs does not exceed the entire capacity of the MEC server since resources have $f_j^{max}$ resource constraints. Constraint C6 specifies that one MD can offload to one MEC server at the most. With this constraint, the objective function based on variable $x_{ij}$ aims to reduce the energy of task offloading. If n tasks exist, then the edge and local processes have $2^{n+1}$ choices. The existence of binary variable $x_{ij} $  changes the optimization problem to a mixed integer programming problem which is non-convex and NP-hard \cite{Lyu2017a, Dai2018a, Chen2018b}. In consideration of problem conditions and different factors, such as the deadline, energy consumption, distance of MDs, and lack of user awareness of others’ decisions, the problem is solved via the school choice.

\begin{figure}[t]
\centering
\includegraphics[width=9cm]{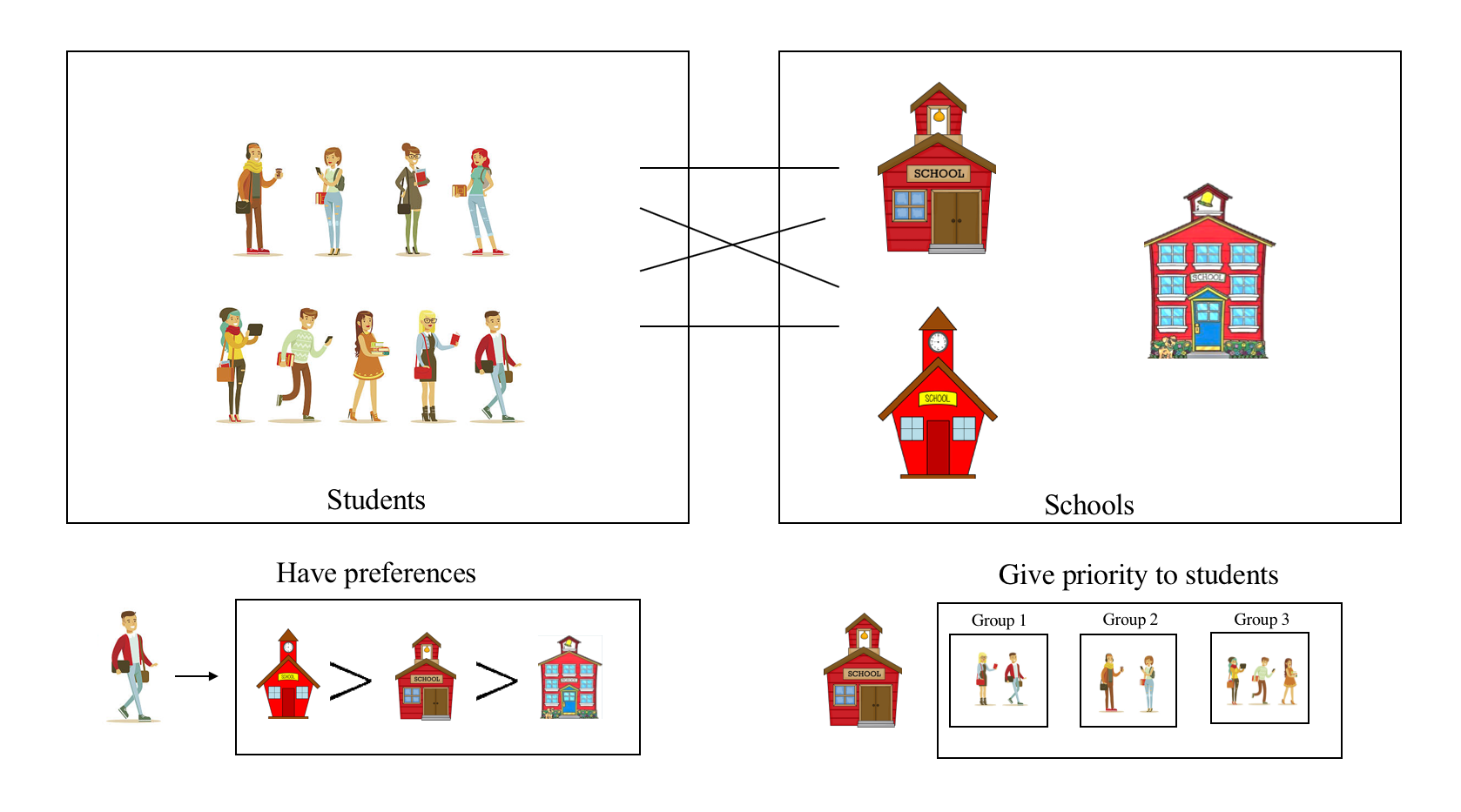}
\caption{School choice}
\label{fig : SchoolChoice}
\end{figure}

\section{Design Of School Choice}

In the following, a new multi-user and multi-MEC scenario is presented where MDs have different requirements. With considering the energy consumption and price, MDs offload their tasks to MEC servers. In a distributed manner, MDs determine task deadlines and workload whereas multi-available MEC servers broadcast their resources to heterogeneous MDs in their area. After that, MDs find available servers and send a request to the best MEC server based on price and energy consumption while each MEC server $j$ has a priority over users. This process continues until all users are assigned or all servers have consumed their resources. The contributions of the present paper are minimizing the energy consumption and price of task execution with consideration of the task deadline and limited capacity in MEC servers. The problem is formulated as a mixed integer programming problem which is non-convex and NP-hard. Therefore, the optimization problem is modeled as a school choice. 

\subsection{What Is the School Choice}

An important application of mechanism design is the school choice \cite{Approach2003a}. Some students are assigned to a number of schools which have a certain capacity with a strict priority over all students.

In this model, schools do not have preferences. In other words, schools do not keep track of which students they register. However, there are specific priorities, such as the distance to the school or a sibling enrolled in the same school. Each student has strict preferences over all schools and an outside option (e.g., attending a private school or being homeschooled). The school choice is an assignment with agents and objects that assigns each student to a school or his/her outside option while respecting the school's capacity and priority.

\subsection{Modeling the Problem as a School Choice}

The current study utilizes the school choice to model the optimization problem, in which MEC operators with resources (modeled as schools) aim to select MDs with better benefits (modeled as students). Accordingly, a decentralized approach is investigated to determine the offloading decision of MDs. With the proposed computation offloading scheme, (i) MDs decide to offload their computation tasks only if offloading to the MEC server satisfies them; (ii) MEC operators choose MDs that they can support; (iii) MDs and servers make their offloading decision in a distributed manner; and (iv) the number of successful offloading tasks increases. 

As shown in Fig. \ref{fig : SchoolChoice}, according to the school's capacity and the student's preference, school choice is a many-to-one assignment in which set of student map to set of schools.  Assigning MDs to MEC servers resembles the problem of school selection by students because many students assign to schools with limited capacity and schools are unable to enroll more students than their capacity. Similarly, many MDs use MEC resources and MEC servers have a finite amount of resources. Consequently, this is also comparable to MEC servers with limited capacity. Within the MDs’ area, servers are located at different distances from MDs and may belong to different operators. Depending on the limitation of the computational resources, the operators can only cover a specified number of MDs.

The school choice consists of players that are agents and objects. In the current paper, MEC servers more closely resemble objects than agents since they offer a service to be used. Therefore, in the present context, MEC servers are assumed to be objects and have a priority over MDs. Also, MDs are selfish and determine task deadlines and workload. Besides, MDs know nothing about other MDs. As a result, as agents, MDs must make the best decisions for offloading and selecting operators. On the other hand, each MEC server considers the interactions among multiple MDs and may either accept or reject MDs based on priorities. In this situation, the school choice is used to deal with the offloading task in multi-user and multi-MEC server scenarios as the short deadline and the speed and efficiency of the offloading tasks are more important than stability. Furthermore, MEC servers have no preference for MDs. In the proposed method, the school choice includes the properties specified below \cite{Haeringera}:

\begin{itemize} 
\item {A set of MDs $U = \{u_1, u_2,…, u_N\} $ }
\item {A set of MEC servers $S = \{s_1, s_2,..., s_M\}$}
\item {A capacity vector $q = (q_{1}.,...,q_{M})$, which specifies for each MEC server $j$, the maximum number of MDs that MEC server can enroll.}
\item {A profile of strict MDs preferences $P = (P_{1},.. .,P_{N})$}
\item {A strict priority structure of the MEC server over MDs, $L = (L_{1},...,L_{M})$}
\item {$U$ and $S$ are kept constant during the decision-making period.}
\item {Each MD $i$ can offload task on one of the MEC servers or perform it locally.}
\item {The MDs and MEC servers are completely independent.}
\end{itemize}

Formally, an assignment is defined as a mapping, $\mu$, from the set of MEC servers and MDs, $ S \cup U$, to a set of all possible sets of MDs and MEC servers:

\begin{equation}
\label{eq:Mapping}
\begin{multlined}
\mu : S \cup U \to 2 ^U \cup S
\end{multlined}
\end{equation}

In the current work, a school choice consists of a population of MDs and a list of MEC servers. MDs are defined by their preferences on MECs, which are the minimum energy consumption for offloading tasks and the price paid to operators. On the other hand, MEC servers are defined by their capacities and priorities in MD selection. Since they have no preferences over MDs and obtain a fixed price based on workloads, MEC servers are not concerned about which MDs send a request. As a result, MEC servers select MDs that are closer to them because of the edge’s limited capacity and the ability to provide better service.

\subsection{Decentralized Solution}

Several approaches can solve the school choice problem, including the Deferred Acceptance Algorithm (DA) and the Immediate Acceptance Algorithm (IA). The IA is very similar to the DA, but with the difference that, once a person occupies a place, this acceptance is final and does not change. As a result, the allocation is performed more rapidly. Furthermore, the immediate acceptance mechanism assigns MDs to MEC servers according to the MD's preferences over MEC servers and the MEC's priorities for MDs. More precisely, at every step $r$, each MEC server accepts (up to its remaining capacity) the highest priority MDs among those that ranked it. 

Therefore, because MECs choose users in their area, a strict priority structure $L_{ij}$ of MEC server $j$ for choosing MD $i$ is only the MD’s distance. 
\begin{equation}
\label{eq:priority}
\begin{multlined}
L_{ij} = D_{ij}
\end{multlined}
\end{equation}

The preference $P_{ij}$ for each MEC server $j$ and MD $i$ is described as follows:

\begin{equation}
\label{eq:preference}
\begin{multlined}
P_{ij} = (\alpha_{i}  * \lambda_j{d_i} ) +  (\beta_{i} *  E_i)
\end{multlined}
\end{equation}

Therefore, the minimum energy consumption for offloading tasks and the payment to operators are the preference $P_{ij}$ for each MEC server $j$ and MD $i$.

The immediate acceptance mechanism is a member of the family of so-called rank-priority mechanisms. Each rank-priority mechanism is associated with an order of all pairs that consists of a (student’s) preference and a (school’s) priority. Given the student’s preferences on schools and the school’s priorities for students, a rank-priority mechanism assigns students to schools following the order of rank-priority pairs. In the IA, the output is efficient. The definition of efficiency is provided as follows \cite{Haeringera}:

%
%
%
%

The assignment is efficient if there is no other assignment in which the number of MDs for at least one MEC server is higher while the other MECs accept fewer MDs. In the current work, the mobile identifies the energy consumption of all servers and first selects the server with the least energy and at a reasonable price. As a result, there is no better server to offload at each stage and each user performs its best and moves on to the next server, only if the preferred server no longer has the capacity to run. Therefore, with the IA, an efficient and quick assignment is achieved within the minimum number of rounds and the lowest response time, which is very practical for MECs and MDs. In the proposed method, the IA in the school choice works as follows:

\begin{itemize} 

\item {MDs find preferences for MEC servers based on the minimum energy consumption and price of MEC servers for offloading tasks in their area.} 
\item {MEC servers set the priority of MDs based on the distance of MDs to MEC servers.} 
\item {MDs apply to their first choice between MEC servers.} 
\item {MEC servers reject the lowest-ranking MDs over their capacity. All other offers are immediately accepted and become permanent assignments. After that, MEC servers capacities are updated again.} 
\item {MDs apply to the next MEC server on their preference list if they rejected in the previous step.}
\item {Eventually, if their lists are empty, MDs apply to the local computing.} 
\item {The algorithm stops when MEC servers use their entire capacity or no MD left without resource.}

\end{itemize}

Based on the fixed MD priorities and submitted preferences, the MD assignment is determined in several rounds. In round one, only the first choice of each MD is considered. Then, one at a time, MDs are assigned to their first choice following their priority order until there is no capacity left at their first choice. In round two, those who cannot be assigned according to their first choice (because of low priority) are considered in regard to their second choice (in case any capacity remains from round one). The process continues in this manner until each MD is assigned \cite{Ergin2003}.

Note that the applications in the IA are immediately accepted or rejected at each step. However, in the DA, acceptance is deferred until the end. This school choice finishes when edge servers have used up all of their capacity and the other MDs either remain unassigned or have all received a resource. In the proposed method, when MEC servers have depleted their capacity and there are still MDs without resources, the MD's decision is to run tasks locally. Thus, when there are no longer MDs without a decision, the algorithm is completed. With the proposed model, by minimizing price and energy consumption before the task deadlines, the offloading decision can be met. In comparison to the DA and game theory, the significant advantage of this schema is the efficient algorithm in the minimum rounds.

Moreover, firstly in the proposed system, users can make offloading decisions to execute the task before the deadline and so save mobile device energy. Secondly, users make these decisions themselves by employing the intelligence on the device. Thirdly, in decentralized offloading, it is unnecessary to send a lot of information for each offloading to the edge. Furthermore, the edge load is reduced and edge management becomes more manageable.

\begin{table}[!t]
\renewcommand{\arraystretch}{1.2} 
\caption{Application Properties }
\label{Application_Properties}
\centering
\begin{tabular}{|c|c|c|c|} 
\hline
\textbf{App name} & \textbf{Data upload(KB)}  & \textbf{Task length(MI)}\\
\hline
Augmented reality  & 1500  & 12000\\
\hline
Health application   & 200 & 6000\\
\hline
Infotainment application  & 250 & 15000\\
\hline
\end{tabular}
\end{table}

\section{Evaluation}
In this section, the EdgeCloudSim simulator evaluates the performance of efficient multi-user and multi-MEC computation offloading. EdgeCloudSim builds upon CloudSim to address the specific demands of edge computing research and support necessary functionality in terms of computation and networking abilities \cite{Sonmez2017}.

\subsection{Parameter Settings}

Through numerical studies, this section evaluates the proposed distributed computation offloading algorithm. To evaluate its algorithm, the current paper uses the following simulation settings for all simulations. First, a scenario is considered in which five MEC servers, with one host and various numbers of computational resources, are deployed in the user's area to serve a different number of users.  The mobile devices have computation-intensive and data-intensive tasks. It is also assumed that the mobile devices are randomly scattered over the coverage region. Thus, the mobile devices can offload tasks to the MEC server. The present study also sets the transmission bandwidth $B_{ij}$ and transmitting power $P_i$ of the mobile device to $20 MHz$ and $0.5 W$, respectively. The channel gain $H_{ij}$ for MD $i$ and MEC server $j$ is modeled as $127 +30 \log_{2}(D_{ij})$, where $D_{ij}$ is the distance between user $i$ and MEC server $j$; this can differ based on the user location. Table \ref{Application_Properties} presents three task types of which each has a different value for input size, file output size, and task length. These properties are generated by an exponential number generator for each task \cite{Sonmez2017}.

\subsection{Use Cases}
In this subsection, we focus on some of the most use cases that experience the need for offloading in mobile edge computing. As shown in table  \ref{Application_Properties}, with respect to our goal, we will take the examples of augmented reality, health application and infotainment application.

\begin{enumerate}
\item Augmented reality is a type of interactive, reality-based display environment that takes the capabilities of effects to increase the user's real-world experience. Augmented reality combines real and computer-based scenes and images to provide a view of the world.

\item One of the major applications of biomedical research is to utilize inexpensive mobile biomedical sensors and cloud computing for pervasive health monitoring. However, real-world user experiences with mobile cloud-based health monitoring were poor, due to factors such as excessive networking latency and longer response time \cite{Cao2015}. So this program can be one of the best programs to take advantage of offloading at the edge.

\item Infotainment app like vehicular infotainment systems includes two parts: 1) the information part and 2) the entertainment part, which are integrated to provide a unified platform to drivers and passengers \cite{Guo2017}. 

\end{enumerate}

\subsection{Performance Evaluation of Algorithm}

The distributed computation offloading algorithm is compared with the following solutions: 

\begin{enumerate}
\item Energy based offloading: In this method, all users offload their computation tasks to the MEC servers that have the lowest energy consumption.
\item Price based offloading: All users offload their computation tasks to the MEC servers that have the lowest price for the users. 
\item HODA: The heuristic offloading decision algorithm (HODA) proposed in \cite{Lyu2017a}.
\end{enumerate}

\begin{figure*}[t!]
\begin{multicols}{3}
    \includegraphics[width=\linewidth]{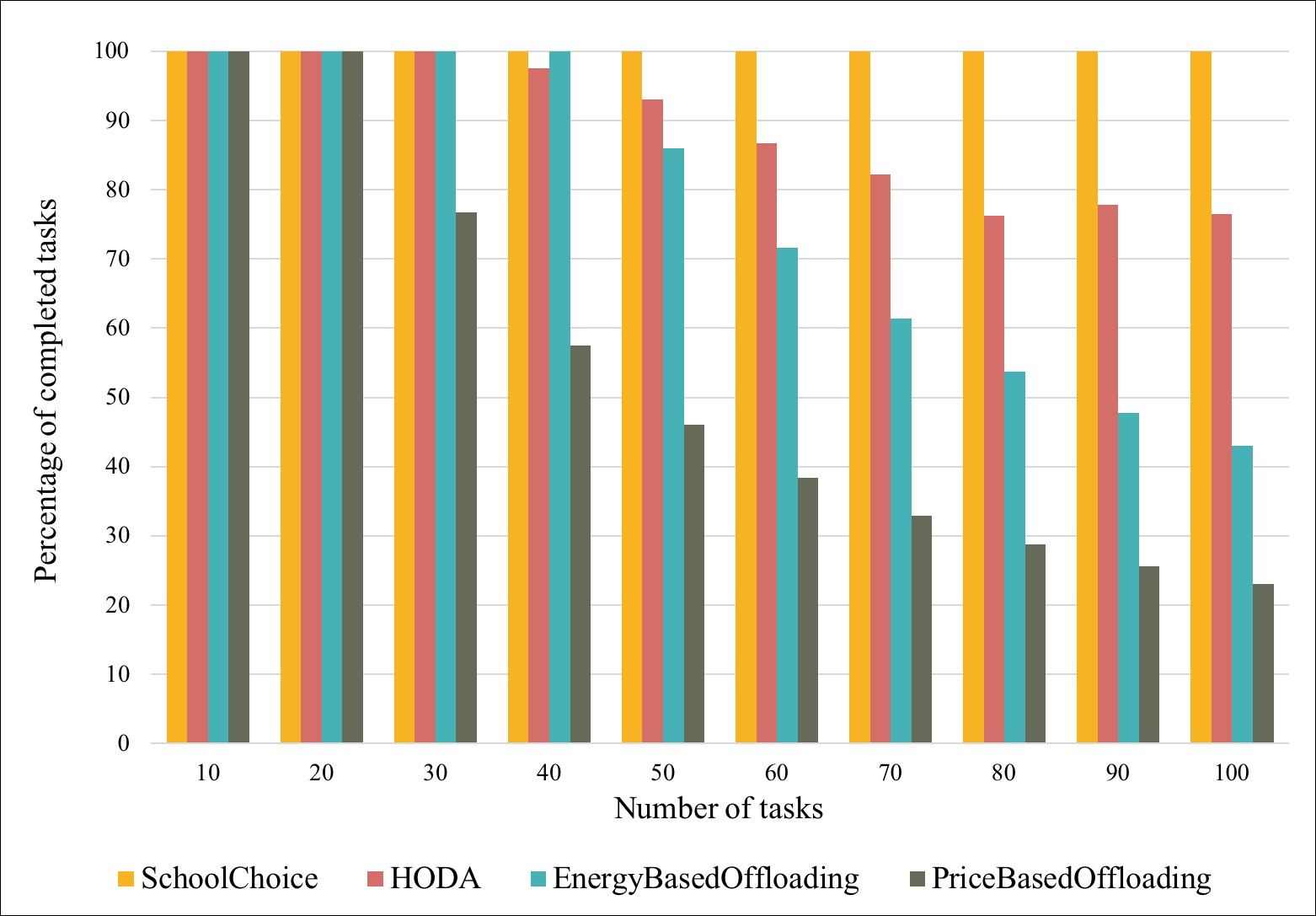}\par 
    \caption{The comparison of algorithms under different numbers of users}
     \label{basic_com.png}
     
    \includegraphics[width=\linewidth]{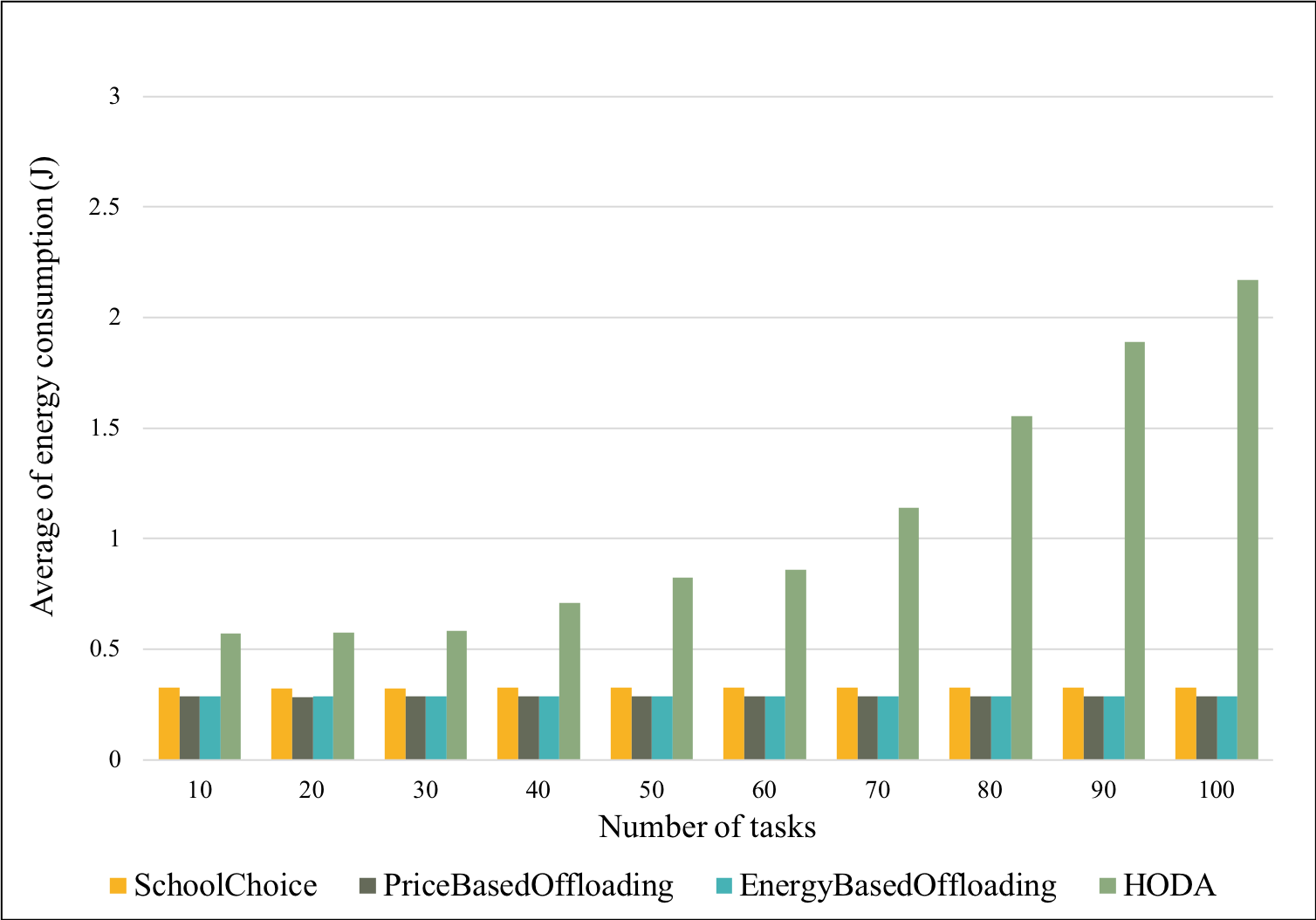}\par  
    \caption{The comparison of energy consumption under different numbers of users}
     \label{basic_energy.png}

    \includegraphics[width=\linewidth]{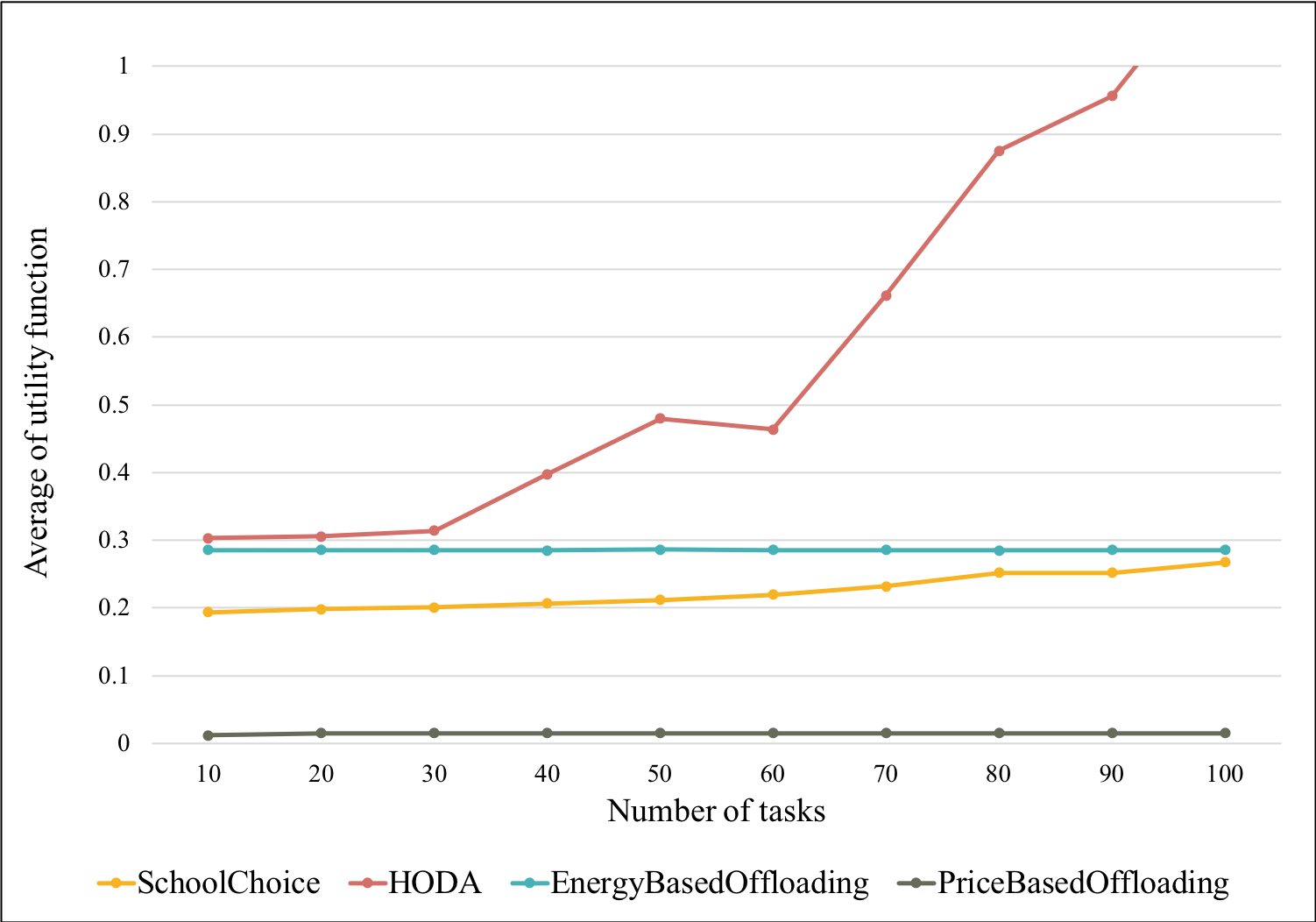}\par
    \caption{the comparison of average utility function under different numbers of users}
          \label{basic_utility.png}

    \end{multicols}
\end{figure*}

The first experiment varies the number of tasks from 10 to 100 and examines the percentage of offloading tasks. In Fig. \ref{basic_com.png}, the percentage of completed tasks is relatively high while the number of tasks is low. However, the percentage of completed tasks gradually decreases as the number of tasks increases. This is reasonable since each user might have a high probability of matching with its preferred MEC server when the number of tasks is low. So they will offload their computation task in good condition, such as bandwidth. Also, as the number of tasks continues to grow, each user must compete with the others for computation resources, thus lowering the probability of a task’s being assigned to its preferred MEC server.  Due to task preferences, the limited number of MEC servers and computational resources, the proposed algorithm must reject some of the user requests and their tasks are executed locally.

It can be seen in Fig. \ref{basic_com.png} that the percentage of users who are able to perform tasks in energy based offloading mode is higher than users in the price based offloading algorithm. Due to features such as distance and workload when users choose based on minimum server energy consumption, the chance to perform tasks before the deadline increased. In this case, if the user fails to fulfill this condition, local execution should be performed. But based on the conditions of this algorithm, local execution is not an option. As a result, the user will not be able to complete the task before the deadline.

Fig. \ref{basic_energy.png} examine the effect of raising the number of tasks on average energy consumption. The present experiment utilizes a health application with five MEC servers. As the number of users increases, users save more energy on school choice scenario than HODA; because HODA does not perform the task allocation well. As a result, the distribution of resources between users is not done correctly. Also, the capacity of edge servers decreases when the number of users increases, and more users are forced to perform their tasks locally. Therefore, the total energy consumption increases due to increased local execution.

Also, in Fig. \ref{basic_utility.png}, the utility function is increased by increasing the number of users and the energy consumption due to more users switching to local execution. On the other hand, in the price-based offloading method, users have less successful execution than the school choice method, because they focus on price parameters which results in a lower utility function than school choice method.

\begin{figure*}[t!]
\begin{multicols}{3}
    \includegraphics[width=\linewidth]{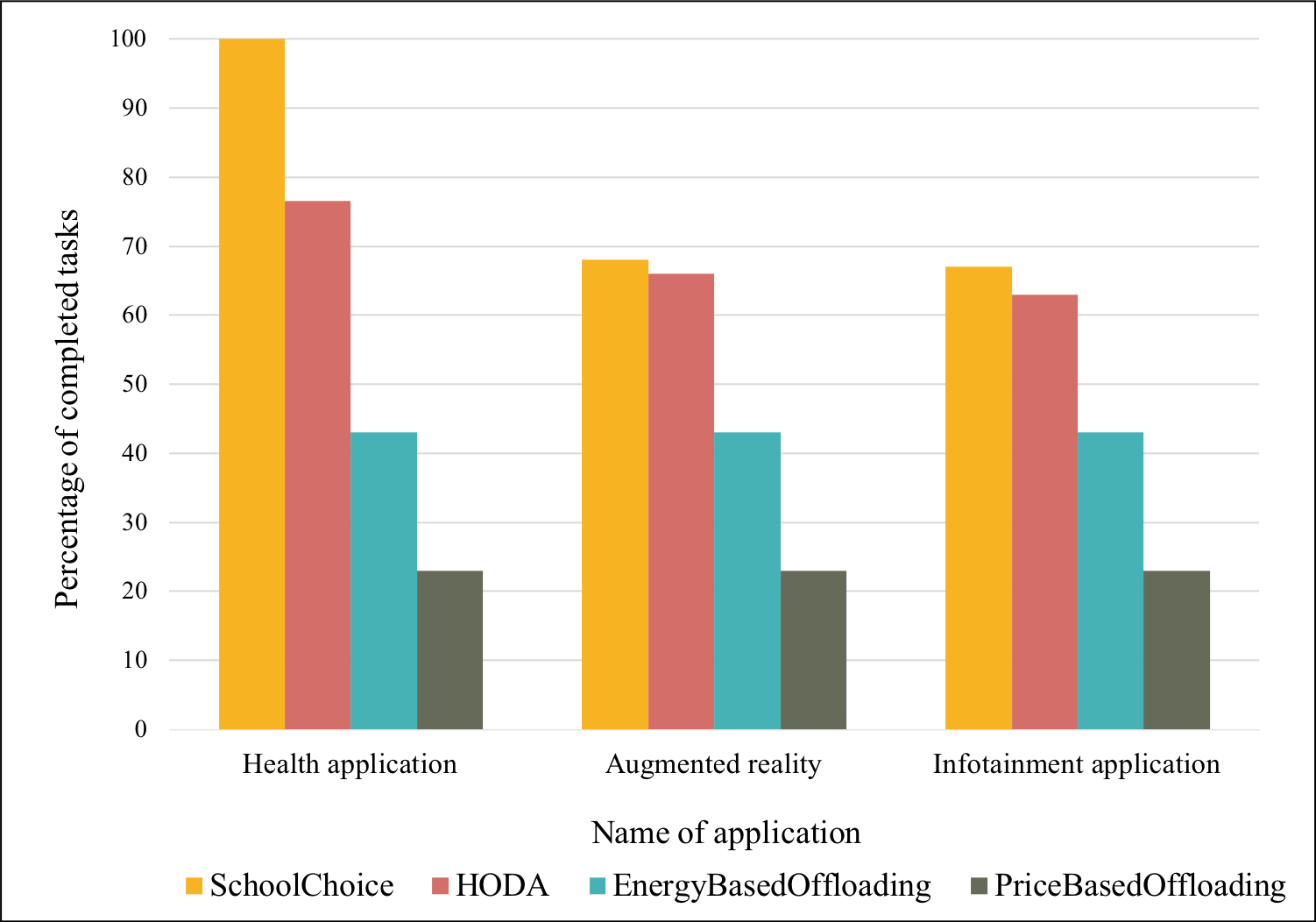}\par 
\caption{The comparison of completed task in different applications}
          \label{app_com.png}

    \includegraphics[width=\linewidth]{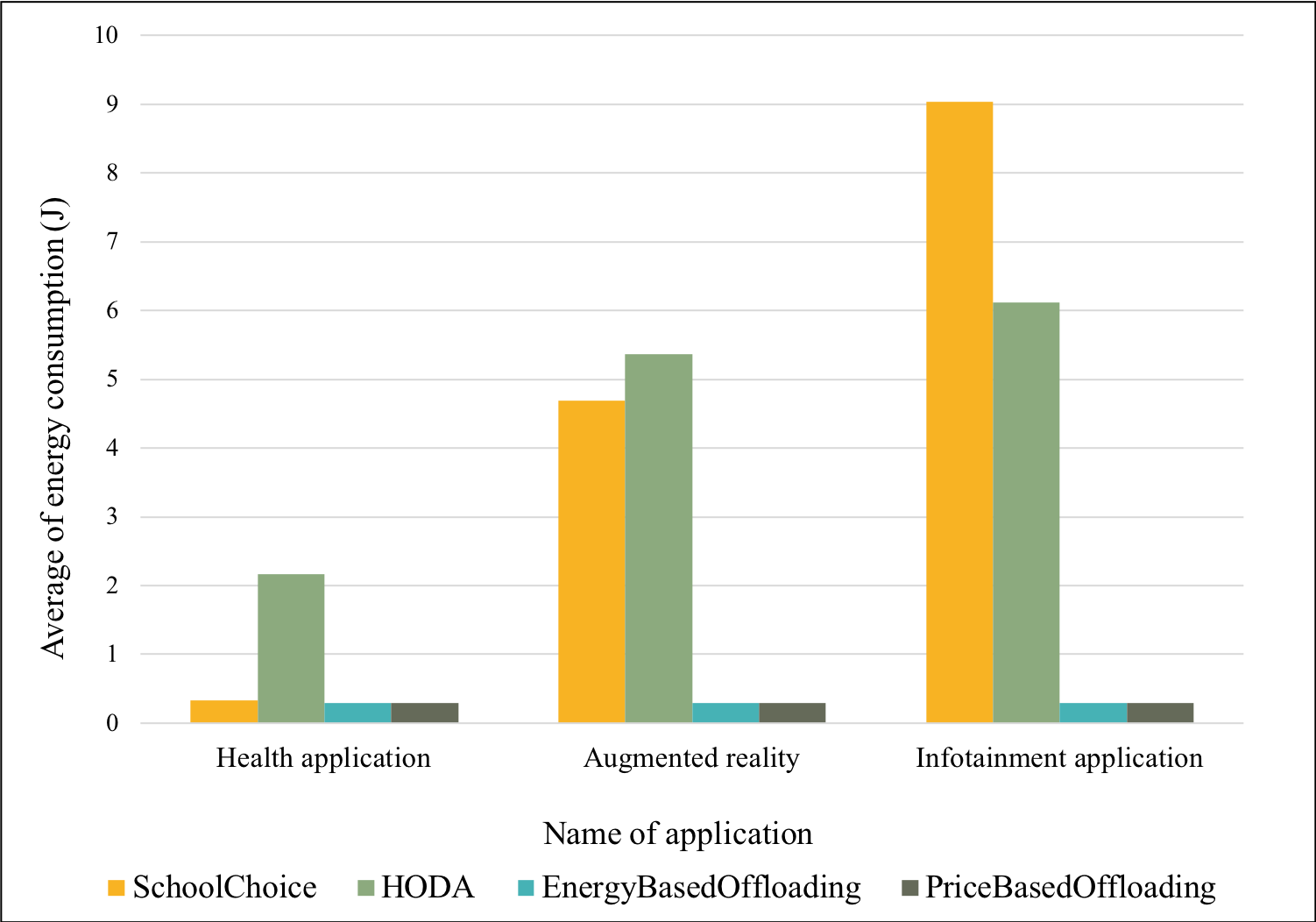}\par  
\caption{The comparison of average energy consumption in different applications}
          \label{app_energy.png}

    \includegraphics[width=\linewidth]{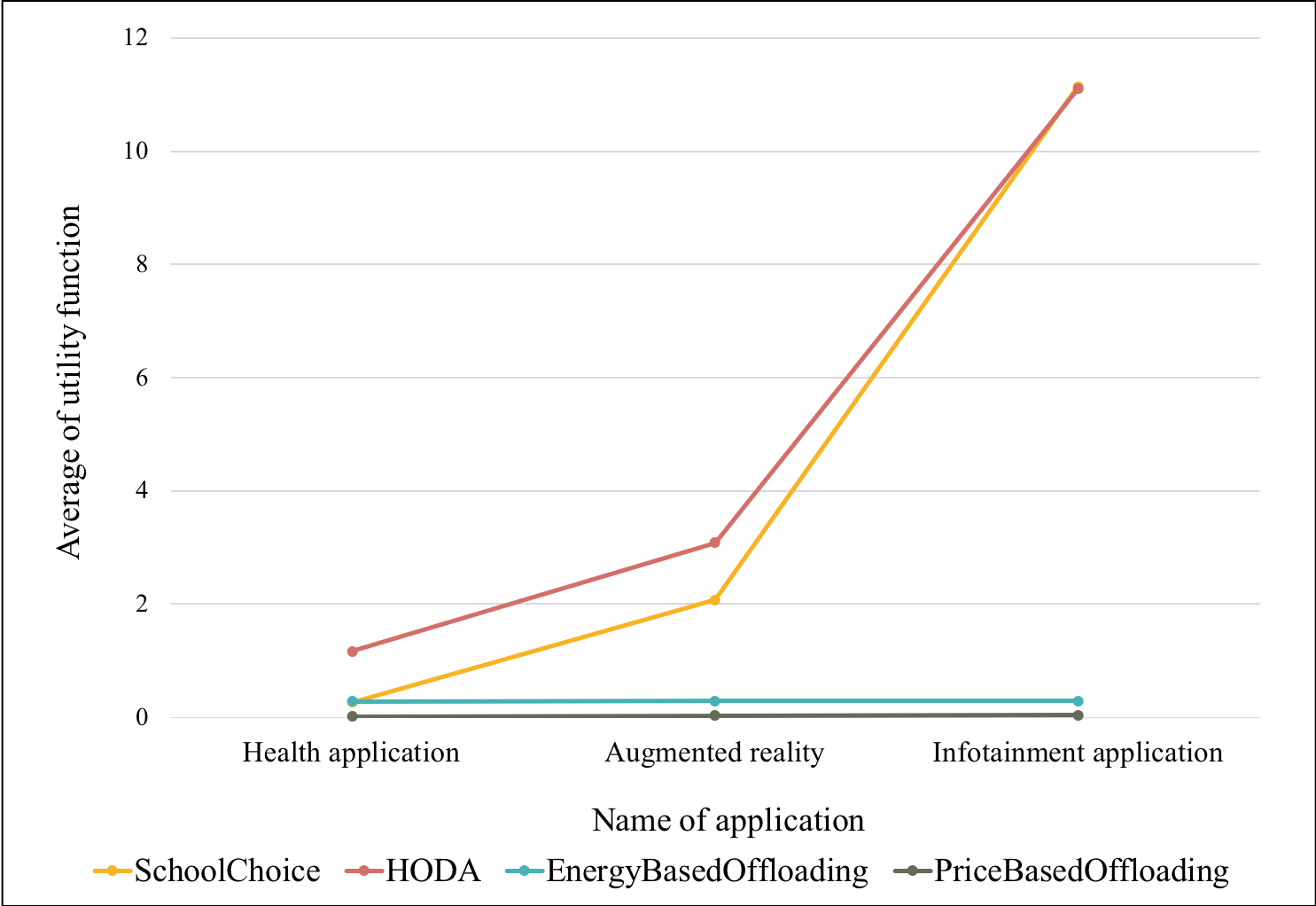}\par
    \caption{The comparison of average utility function in different applications}
              \label{app_utility.png}
     
    \end{multicols}
\end{figure*}

In this paper, multiple heterogeneous MDs compete for numerous heterogeneous MEC resources as the workload increases. Hence, the offloading scenario will be more difficult, and some users will have to select the local execution mode, which may not be able to perform the tasks properly due to the high workload of tasks. As shown in Fig. \ref{app_com.png}, as the workloads increase, the percentage of users who perform their tasks is reduced because the edge server has limited resources and cannot handle all users. On the other hand, in the mobile device, the user will not have the capacity to perform these tasks successfully before the deadline, which will result in unsuccessful execution of tasks. In the price-based offloading method, price is the only important factor for the user, and the workload has a direct impact on the user's payment. The price paid by the user also increases when workloads increment.

In many cases, this amount exceeds the maximum price that the user can pay, and the user is forced to perform locally. In this condition, users can bid the second price through methods such as an auction. However, this article assumed that the maximum user price is only set once at the beginning. Hence, tasks are forced to run locally, but they are unable to perform before the deadline in some cases, due to the increased workload. As shown in Fig. \ref{app_utility.png}, although utility function has improved dramatically with increasing workload, the school choice method has still been less than HODA method due to better allocation based on both energy factor and user price.

\begin{figure*}[t!]
\begin{multicols}{3}
    \includegraphics[width=\linewidth]{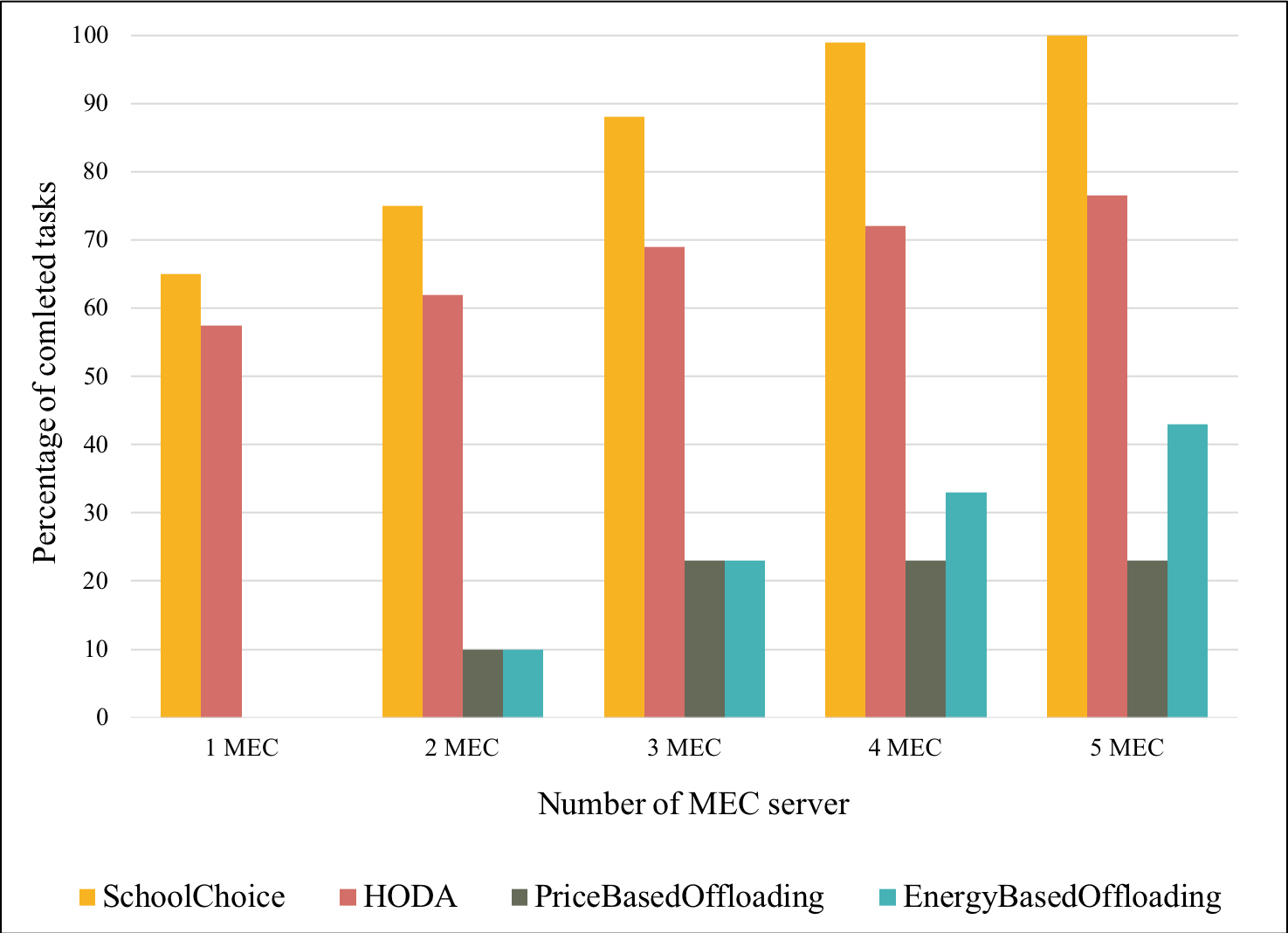}\par 
\caption{The comparison of completed task in different number of MEC servers}
          \label{MEC_com.png}

    \includegraphics[width=\linewidth]{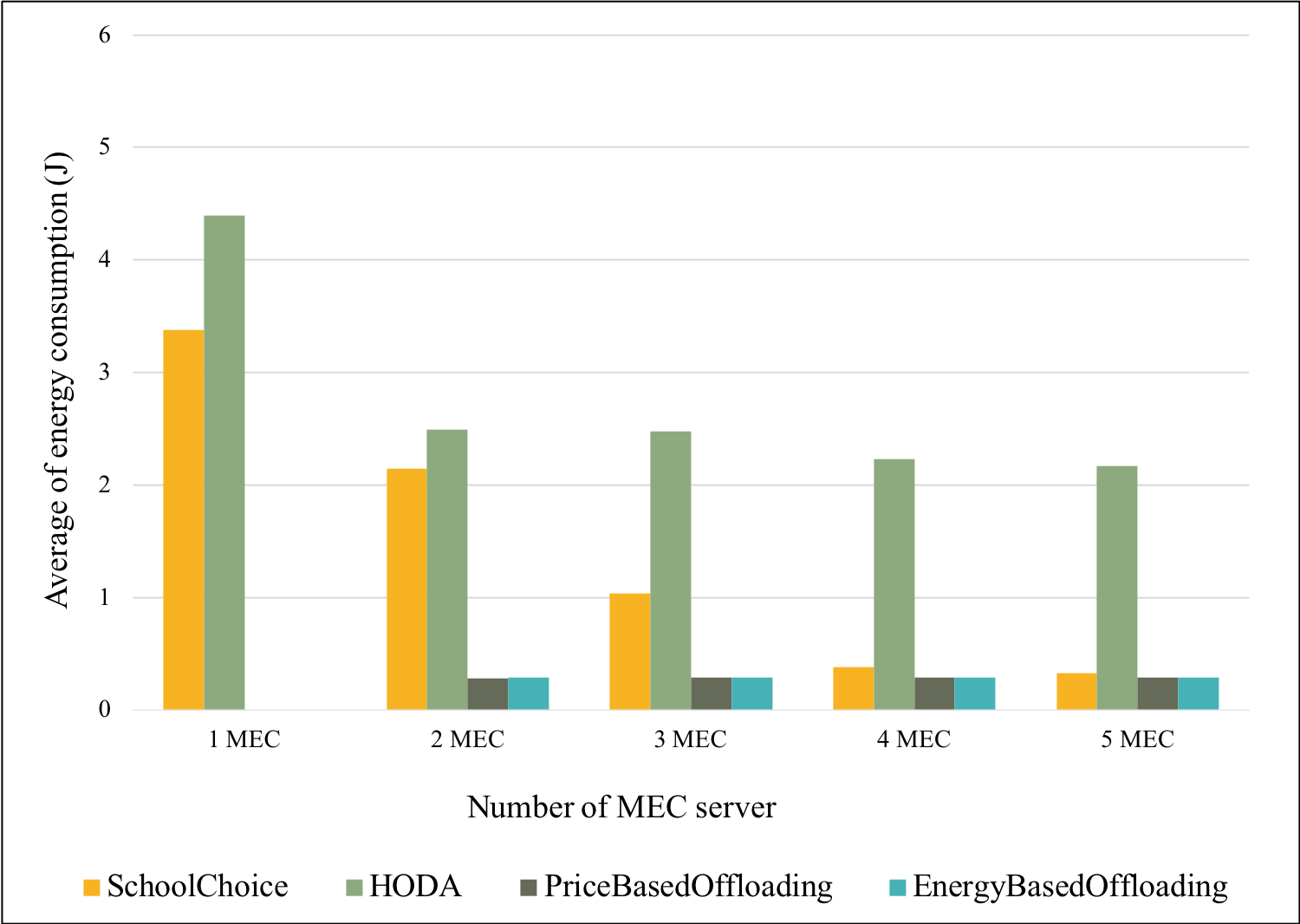}\par  
\caption{The comparison of average energy consumption in different number of MEC servers}
          \label{MEC_energy.png}

    \includegraphics[width=\linewidth]{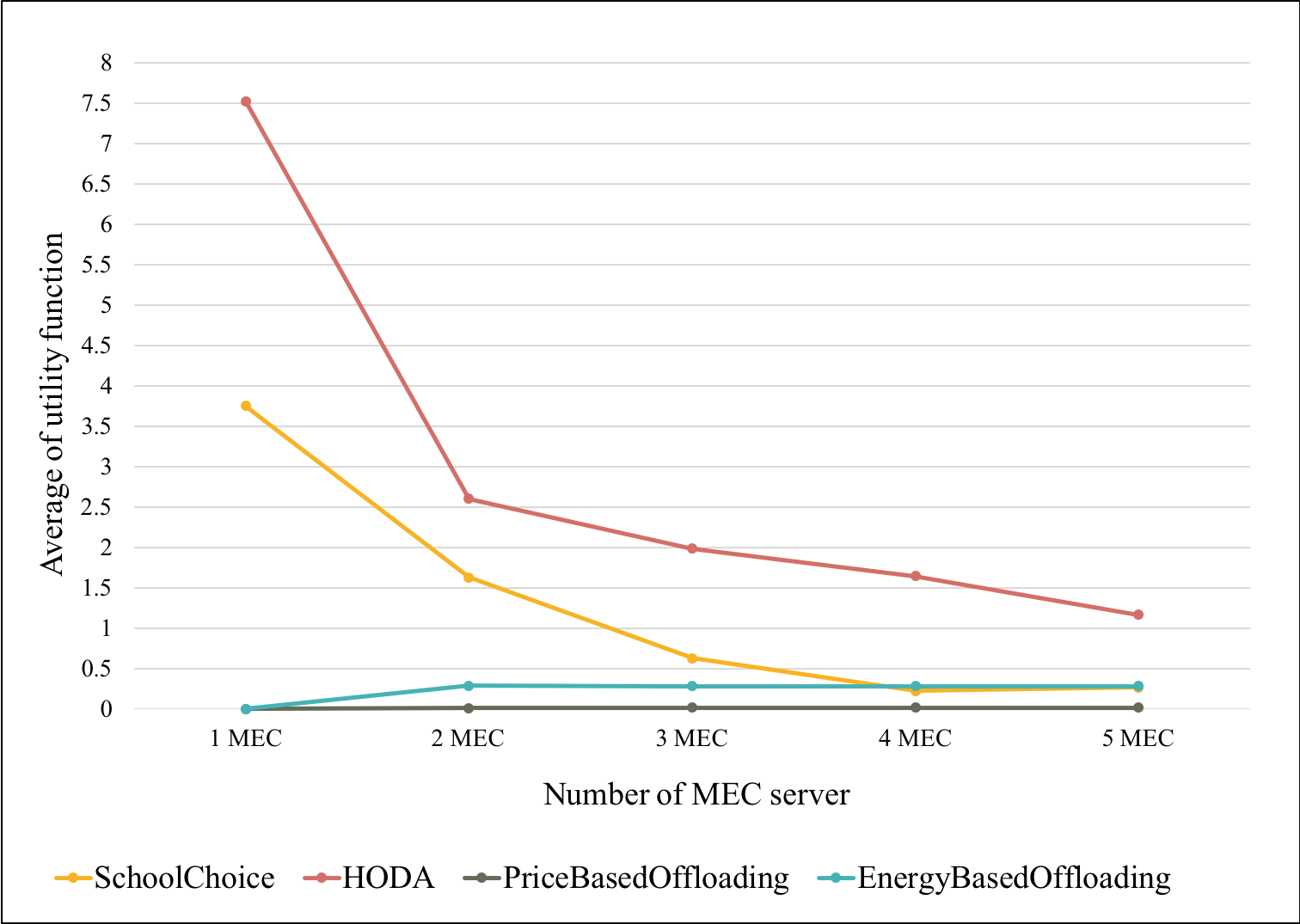}\par
    \caption{The comparison of average utility function in different number of MEC servers}
              \label{MEC_utility.png}

    \end{multicols}
\end{figure*}

In Fig. \ref{MEC_com.png}, with the increase in the number of servers to 5, the school choice approach has reached its best, and all users can offload their tasks to a server with low energy consumption and cost based on user-selected parameters. However, in this case, HODA still has to push some users to run locally, because it cannot determine the appropriate allocation to perform tasks based on the user's parameters. As a result in range of 100 concurrent users, as shown in Fig. \ref{MEC_energy.png}, the average energy consumption increase.

Fig. \ref{MEC_energy.png} also shows that with the increase in MEC server resources and the reduction of resource constraints, people's chances of offloading tasks increase. As a result, the average energy consumption is reduced due to reduced local execution. However, as in the previous cases, the energy consumption, when both energy and price parameters are equal, has a more significant impact on the success of the users, and if the user chooses a server with less energy consumption, the chances of successful execution before the deadline are increased. In Fig. \ref{MEC_utility.png}, as the number of servers increases, the range of suggested base prices are increased. Given all of this, with increasing the chance of successful execution before the deadline, the utility function will also decrease. It is due to the increasing number of edge server resources and the increasing probability of users to find the server at lower price and energy consumption.

\section{Conclusion}

With the aim of reducing mobile device energy consumption and price, the present research studied multi-user and multi-MEC computation offloading. By formulating different task offloading decisions as a school choice problem, the current study minimized user energy consumption and price under latency constraints. Instead of conventional centralized optimization methods, this approach considered a decentralized mechanism between users and MEC servers. Simulation results indicated that the proposed scheme is more efficient in comparison with other computation offloading schemes. Future work shall consider task offloading in more complicated deployment with user mobility.

%
%
%

\section*{References}

\bibliography{RefrenceOfPaper.bib}

\begin{thebibliography}{10}
\expandafter\ifx\csname url\endcsname\relax
  \def\url#1{\texttt{#1}}\fi
\expandafter\ifx\csname urlprefix\endcsname\relax\def\urlprefix{URL }\fi
\expandafter\ifx\csname href\endcsname\relax
  \def\href#1#2{#2} \def\path#1{#1}\fi

\bibitem{Hu2015b}
Y.~C. Hu, M.~Patel, D.~Sabella, N.~Sprecher, V.~Young,
  \href{https://yucianga.info/wp-content/uploads/2015/11/Ref02-2015-09-etsi{\_}wp11{\_}mec{\_}a{\_}key{\_}technology{\_}towards{\_}5g.pdf}{{Mobile
  edge computing—A key technology towards 5G. ETSI White Paper}}, ETSI White
  Pap. 11~(11) (2015) 1--16.
\newline\urlprefix\url{https://yucianga.info/wp-content/uploads/2015/11/Ref02-2015-09-etsi{\_}wp11{\_}mec{\_}a{\_}key{\_}technology{\_}towards{\_}5g.pdf}

\bibitem{Li2018Users}
K.~Li, \href{https://ieeexplore.ieee.org/document/8454762/}{{A Game Theoretic
  Approach to Computation Offloading Strategy Optimization for Non-cooperative
  Users in Mobile Edge Computing}}, IEEE Trans. Sustain. Comput. (2018)
  1--1\href {http://dx.doi.org/10.1109/TSUSC.2018.2868655}
  {\path{doi:10.1109/TSUSC.2018.2868655}}.
\newline\urlprefix\url{https://ieeexplore.ieee.org/document/8454762/}

\bibitem{Zhang2016}
K.~Zhang, Y.~Mao, S.~Leng, Q.~Zhao, L.~Li, X.~Peng, L.~Pan, S.~Maharjan,
  Y.~Zhang, \href{http://www.mdpi.com/2076-3417/7/6/557}{{Energy-Efficient
  Offloading for Mobile Edge Computing in 5G Heterogeneous Networks}}, IEEE
  Access 4~(c) (2016) 5896--5907.
\newblock \href {http://dx.doi.org/10.1109/ACCESS.2016.2597169}
  {\path{doi:10.1109/ACCESS.2016.2597169}}.
\newline\urlprefix\url{http://www.mdpi.com/2076-3417/7/6/557}

\bibitem{Zhao2017}
P.~Zhao, H.~Tian, C.~Qin, G.~Nie, {Energy-Saving Offloading by Jointly
  Allocating Radio and Computational Resources for Mobile Edge Computing}, IEEE
  Access 5 (2017) 11255--11268.
\newblock \href {http://dx.doi.org/10.1109/ACCESS.2017.2710056}
  {\path{doi:10.1109/ACCESS.2017.2710056}}.

\bibitem{Zhang2017a}
J.~Zhang, X.~Hu, Z.~Ning, E.~C. Ngai, L.~Zhou, J.~Wei, J.~Cheng, B.~Hu,
  {Energy-latency Trade-off for Energy-aware Offloading in Mobile Edge
  Computing Networks}, IEEE Internet Things J. 4662~(c) (2017) 1--13.
\newblock \href {http://dx.doi.org/10.1109/JIOT.2017.2786343}
  {\path{doi:10.1109/JIOT.2017.2786343}}.

\bibitem{Hao2018}
Y.~Hao, M.~Chen, L.~Hu, M.~S. Hossain, A.~Ghoneim, {Energy Efficient Task
  Caching and Offloading for Mobile Edge Computing}, IEEE Access 6~(March)
  (2018) 11365--11373.
\newblock \href {http://dx.doi.org/10.1109/ACCESS.2018.2805798}
  {\path{doi:10.1109/ACCESS.2018.2805798}}.

\bibitem{Zhang2018}
G.~Zhang, W.~Zhang, Y.~Cao, D.~Li, L.~Wang,
  \href{https://ieeexplore.ieee.org/document/8371267/}{{Energy-Delay Tradeoff
  for Dynamic Offloading in Mobile-Edge Computing System with Energy Harvesting
  Devices}}, IEEE Trans. Ind. Informatics 3203~(c) (2018) 1--1.
\newblock \href {http://dx.doi.org/10.1109/TII.2018.2843365}
  {\path{doi:10.1109/TII.2018.2843365}}.
\newline\urlprefix\url{https://ieeexplore.ieee.org/document/8371267/}

\bibitem{You2017}
C.~You, K.~Huang, H.~Chae, B.~H. Kim, {Energy-Efficient Resource Allocation for
  Mobile-Edge Computation Offloading}, IEEE Trans. Wirel. Commun. 16~(3) (2017)
  1397--1411.
\newblock \href {http://arxiv.org/abs/1605.08518} {\path{arXiv:1605.08518}},
  \href {http://dx.doi.org/10.1109/TWC.2016.2633522}
  {\path{doi:10.1109/TWC.2016.2633522}}.

\bibitem{Li2019a}
S.~Li, Z.~Zhang, P.~Zhang, X.~Qin, Y.~Tao, L.~Liu, {Energy-aware Mobile Edge
  Computation Offloading for IoT over Heterogenous Networks}, IEEE Access 7
  (2019) 1--1.
\newblock \href {http://dx.doi.org/10.1109/access.2019.2893118}
  {\path{doi:10.1109/access.2019.2893118}}.

\bibitem{8241344}
W.~Fan, Y.~Liu, B.~Tang, F.~Wu, Z.~Wang, {Computation Offloading Based on
  Cooperations of Mobile Edge Computing-Enabled Base Stations}, IEEE Access
  6~(X) (2017) 22622--22633.
\newblock \href {http://dx.doi.org/10.1109/ACCESS.2017.2787737}
  {\path{doi:10.1109/ACCESS.2017.2787737}}.

\bibitem{Guo2018}
F.~Guo, H.~Zhang, H.~Ji, X.~Li, V.~C. Leung, {An Efficient Computation
  Offloading Management Scheme in the Densely Deployed Small Cell Networks With
  Mobile Edge Computing}, IEEE/ACM Trans. Netw. (2018) 1--14\href
  {http://dx.doi.org/10.1109/TNET.2018.2873002}
  {\path{doi:10.1109/TNET.2018.2873002}}.

\bibitem{Dai2018a}
Y.~Dai, D.~Xu, S.~Maharjan, Y.~Zhang, {Joint Computation Offloading and User
  Association in Multi-task Mobile Edge Computing}, IEEE Trans. Veh. Technol.
  9545~(c) (2018) 1--13.
\newblock \href {http://dx.doi.org/10.1109/TVT.2018.2876804}
  {\path{doi:10.1109/TVT.2018.2876804}}.

\bibitem{Tran2017}
T.~X. Tran, D.~Pompili, \href{http://arxiv.org/abs/1705.00704}{{Joint Task
  Offloading and Resource Allocation for Multi-Server Mobile-Edge Computing
  Networks}}, IEEE Access 5 (2017) 3302--3312.
\newblock \href {http://arxiv.org/abs/1705.00704} {\path{arXiv:1705.00704}}.
\newline\urlprefix\url{http://arxiv.org/abs/1705.00704}

\bibitem{Dinh2017}
T.~Q. Dinh, J.~Tang, Q.~D. La, T.~Q. Quek, {Offloading in Mobile Edge
  Computing: Task Allocation and Computational Frequency Scaling}, IEEE Trans.
  Commun. 65~(8) (2017) 3571--3584.
\newblock \href {http://dx.doi.org/10.1109/TCOMM.2017.2699660}
  {\path{doi:10.1109/TCOMM.2017.2699660}}.

\bibitem{Chen2018b}
M.~Chen, Y.~Hao, {Task Offloading for Mobile Edge Computing in Software Defined
  Ultra-Dense Network}, IEEE J. Sel. Areas Commun. 36~(3) (2018) 587--597.
\newblock \href {http://dx.doi.org/10.1109/JSAC.2018.2815360}
  {\path{doi:10.1109/JSAC.2018.2815360}}.

\bibitem{Ugwuanyi2018}
E.~E. Ugwuanyi, S.~Ghosh, M.~Iqbal, T.~Dagiuklas, {Reliable resource
  provisioning using bankers' deadlock avoidance algorithm in MEC for
  industrial IoT}, IEEE Access 6 (2018) 43327--43335.
\newblock \href {http://dx.doi.org/10.1109/ACCESS.2018.2857726}
  {\path{doi:10.1109/ACCESS.2018.2857726}}.

\bibitem{Huang}
L.~Huang, X.~Feng, L.~Zhang, L.~Qian, Y.~Wu,
  \href{https://www.mdpi.com/1424-8220/19/6/1446}{{Multi-Server Multi-User
  Multi-Task Computation Offloading for Mobile Edge Computing Networks}},
  Sensors 19~(6) (2019) 1446.
\newblock \href {http://dx.doi.org/10.3390/s19061446}
  {\path{doi:10.3390/s19061446}}.
\newline\urlprefix\url{https://www.mdpi.com/1424-8220/19/6/1446}

\bibitem{Li2019c}
K.~Li, {Computation Offloading Strategy Optimization with Multiple
  Heterogeneous Servers in Mobile Edge Computing}, IEEE Trans. Sustain. Comput.
  XX (2019) 1--1.
\newblock \href {http://dx.doi.org/10.1109/tsusc.2019.2904680}
  {\path{doi:10.1109/tsusc.2019.2904680}}.

\bibitem{Nguyen2018}
D.~T. Nguyen, L.~B. Le, V.~Bhargava,
  \href{http://arxiv.org/abs/1805.02982}{{Price-based Resource Allocation for
  Edge Computing: A Market Equilibrium Approach}}, IEEE Trans. Cloud Comput.
  (2018) 1--19\href {http://arxiv.org/abs/1805.02982}
  {\path{arXiv:1805.02982}}, \href {http://dx.doi.org/10.1109/TCC.2018.2844379}
  {\path{doi:10.1109/TCC.2018.2844379}}.
\newline\urlprefix\url{http://arxiv.org/abs/1805.02982}

\bibitem{Gao2019}
G.~Gao, M.~Xiao, J.~Wu, H.~Huang, S.~Wang, G.~Chen,
  \href{https://ieeexplore.ieee.org/document/8657771/}{{Auction-based VM
  Allocation for Deadline-Sensitive Tasks in Distributed Edge Cloud}}, IEEE
  Trans. Serv. Comput. PP~(201806340014) (2019) 1--1.
\newblock \href {http://dx.doi.org/10.1109/TSC.2019.2902549}
  {\path{doi:10.1109/TSC.2019.2902549}}.
\newline\urlprefix\url{https://ieeexplore.ieee.org/document/8657771/}

\bibitem{8166725}
M.~Liu, Y.~Liu,
  \href{https://ieeexplore.ieee.org/abstract/document/8166725/}{{Price-Based
  Distributed Offloading for Mobile-Edge Computing with Computation Capacity
  Constraints}}, IEEE Wirel. Commun. Lett. 7~(3) (2018) 420--423.
\newblock \href {http://arxiv.org/abs/1712.00599} {\path{arXiv:1712.00599}},
  \href {http://dx.doi.org/10.1109/LWC.2017.2780128}
  {\path{doi:10.1109/LWC.2017.2780128}}.
\newline\urlprefix\url{https://ieeexplore.ieee.org/abstract/document/8166725/}

\bibitem{Ranadheera2018}
S.~Ranadheera, S.~Maghsudi, E.~Hossain, {Computation Offloading and Activation
  of Mobile Edge Computing Servers: A Minority Game}, IEEE Wirel. Commun. Lett.
  (2018) 1--4\href {http://arxiv.org/abs/1710.05499} {\path{arXiv:1710.05499}},
  \href {http://dx.doi.org/10.1109/LWC.2018.2810292}
  {\path{doi:10.1109/LWC.2018.2810292}}.

\bibitem{Li2018f}
M.~Li, Q.~Wu, J.~Zhu, R.~Zheng, M.~Zhang,
  \href{https://www.hindawi.com/journals/wcmc/2018/2179316/}{{A Computing
  Offloading Game for Mobile Devices and Edge Cloud Servers}}, Wirel. Commun.
  Mob. Comput. 2018 (2018) 1--10.
\newblock \href {http://dx.doi.org/10.1155/2018/2179316}
  {\path{doi:10.1155/2018/2179316}}.
\newline\urlprefix\url{https://www.hindawi.com/journals/wcmc/2018/2179316/}

\bibitem{Zhang2017d}
H.~Zhang, F.~Guo, H.~Ji, C.~Zhu, {Combinational auction-based service provider
  selection in mobile edge computing networks}, IEEE Access 5 (2017)
  13455--13464.
\newblock \href {http://dx.doi.org/10.1109/ACCESS.2017.2721957}
  {\path{doi:10.1109/ACCESS.2017.2721957}}.

\bibitem{Gu2018}
Y.~Gu, Z.~Chang, M.~Pan, L.~Song, Z.~Han, {Joint Radio and Computational
  Resource Allocation in IoT Fog Computing}, IEEE Trans. Veh. Technol. 67~(8)
  (2018) 7475--7484.
\newblock \href {http://dx.doi.org/10.1109/TVT.2018.2820838}
  {\path{doi:10.1109/TVT.2018.2820838}}.

\bibitem{Gu2018a}
B.~Gu, Y.~Chen, H.~Liao, Z.~Zhou, D.~Zhang, {A distributed and context-aware
  task assignment mechanism for collaborative mobile edge computing}, Sensors
  (Switzerland) 18~(8).
\newblock \href {http://dx.doi.org/10.3390/s18082423}
  {\path{doi:10.3390/s18082423}}.

\bibitem{Ren2018}
J.~Ren, G.~Yu, Y.~Cai, Y.~He, \href{http://arxiv.org/abs/1704.00163}{{Latency
  optimization for resource allocation in mobile-edge computation offloading}},
  IEEE Trans. Wirel. Commun. 17~(8) (2018) 5506--5519.
\newblock \href {http://arxiv.org/abs/1704.00163} {\path{arXiv:1704.00163}},
  \href {http://dx.doi.org/10.1109/TWC.2018.2845360}
  {\path{doi:10.1109/TWC.2018.2845360}}.
\newline\urlprefix\url{http://arxiv.org/abs/1704.00163}

\bibitem{Yang2018}
L.~Yang, H.~Zhang, X.~Li, H.~Ji, V.~C. Leung, {A Distributed Computation
  Offloading Strategy in Small-Cell Networks Integrated With Mobile Edge
  Computing}, IEEE/ACM Trans. Netw. (2018) 1--12\href
  {http://dx.doi.org/10.1109/TNET.2018.2876941}
  {\path{doi:10.1109/TNET.2018.2876941}}.

\bibitem{Dinh2018}
T.~Q. Dinh, Q.~D. La, T.~Q.~S. Quek, H.~Shin,
  \href{https://ieeexplore.ieee.org/document/8444467/}{{Distributed Learning
  for Computation Offloading in Mobile Edge Computing}}, IEEE Trans. Commun.
  66~(12) (2018) 6353--6367.
\newblock \href {http://dx.doi.org/10.1109/TCOMM.2018.2866572}
  {\path{doi:10.1109/TCOMM.2018.2866572}}.
\newline\urlprefix\url{https://ieeexplore.ieee.org/document/8444467/}

\bibitem{Bahreini2018}
T.~Bahreini, D.~Grosu, {An Envy-Free Auction Mechanism for Resource Allocation
  in Edge Computing Systems}, 2018 IEEE/ACM Symp. Edge Comput. (2018)
  313--322\href {http://dx.doi.org/10.1109/SEC.2018.00030}
  {\path{doi:10.1109/SEC.2018.00030}}.

\bibitem{Sun2018}
W.~Sun, J.~Liu, Y.~Yue, H.~Zhang,
  \href{https://ieeexplore.ieee.org/document/8410767/}{{Double Auction-based
  Resource Allocation for Mobile Edge Computing in Industrial Internet of
  Things}}, IEEE Trans. Ind. Informatics PP~(c) (2018) 1.
\newblock \href {http://dx.doi.org/10.1109/TII.2018.2855746}
  {\path{doi:10.1109/TII.2018.2855746}}.
\newline\urlprefix\url{https://ieeexplore.ieee.org/document/8410767/}

\bibitem{Chen2016}
X.~Chen, L.~Jiao, W.~Li, X.~Fu,
  \href{https://ieeexplore.ieee.org/iel7/90/4359146/07307234.pdf}{{Efficient
  Multi-User Computation Offloading for Mobile-Edge Cloud Computing}}, IEEE/ACM
  Trans. Netw. 24~(5) (2016) 2795--2808.
\newblock \href {http://arxiv.org/abs/1510.00888} {\path{arXiv:1510.00888}},
  \href {http://dx.doi.org/10.1109/TNET.2015.2487344}
  {\path{doi:10.1109/TNET.2015.2487344}}.
\newline\urlprefix\url{https://ieeexplore.ieee.org/iel7/90/4359146/07307234.pdf}

\bibitem{Guo2018b}
H.~Guo, J.~Liu, J.~Zhang, W.~Sun, N.~Kato,
  \href{https://ieeexplore.ieee.org/abstract/document/8361406/
  https://ieeexplore.ieee.org/document/8361406/}{{Mobile-Edge Computation
  Offloading for Ultra-Dense IoT Networks}}, IEEE Internet Things J. (2018)
  1\href {http://dx.doi.org/10.1109/JIOT.2018.2838584}
  {\path{doi:10.1109/JIOT.2018.2838584}}.
\newline\urlprefix\url{https://ieeexplore.ieee.org/abstract/document/8361406/
  https://ieeexplore.ieee.org/document/8361406/}

\bibitem{Yang2018cooperative}
B.~Yang, Z.~Li, W.~Liu, \href{http://arxiv.org/abs/1812.07781}{{Non-cooperative
  game approach for task offloading in edge clouds}}, arXiv (2018) 1--12\href
  {http://arxiv.org/abs/1812.07781} {\path{arXiv:1812.07781}}.
\newline\urlprefix\url{http://arxiv.org/abs/1812.07781}

\bibitem{Li2018}
N.~Li, J.~F. Martinez-Ortega, V.~H. Diaz, \href{http://arxiv.org/abs/1805.02182
  https://ieeexplore.ieee.org/document/8390908/}{{Distributed power control for
  interference-aware multi-user mobile edge computing: A game theory
  approach}}, IEEE Access 6~(i) (2018) 36105--36114.
\newblock \href {http://arxiv.org/abs/1805.02182} {\path{arXiv:1805.02182}},
  \href {http://dx.doi.org/10.1109/ACCESS.2018.2849207}
  {\path{doi:10.1109/ACCESS.2018.2849207}}.
\newline\urlprefix\url{http://arxiv.org/abs/1805.02182
  https://ieeexplore.ieee.org/document/8390908/}

\bibitem{Yi2019}
C.~Yi, J.~Cai, Z.~Su, \href{https://ieeexplore.ieee.org/document/8606230/}{{A
  Multi-User Mobile Computation Offloading and Transmission Scheduling
  Mechanism for Delay-Sensitive Applications}}, IEEE Trans. Mob. Comput. (2019)
  1--1\href {http://dx.doi.org/10.1109/TMC.2019.2891736}
  {\path{doi:10.1109/TMC.2019.2891736}}.
\newline\urlprefix\url{https://ieeexplore.ieee.org/document/8606230/}

\bibitem{Zhang2017}
T.~Zhang, \href{http://ieeexplore.ieee.org/document/8226751/
  http://arxiv.org/abs/1709.04148}{{Data Offloading in Mobile Edge Computing: A
  Coalition and Pricing Based Approach}}, IEEE Access 6 (2018) 2760--2767.
\newblock \href {http://arxiv.org/abs/1709.04148} {\path{arXiv:1709.04148}},
  \href {http://dx.doi.org/10.1109/ACCESS.2017.2785265}
  {\path{doi:10.1109/ACCESS.2017.2785265}}.
\newline\urlprefix\url{http://ieeexplore.ieee.org/document/8226751/
  http://arxiv.org/abs/1709.04148}

\bibitem{8249785}
Z.~Zhu, J.~Peng, X.~Gu, H.~Li, K.~Liu, Z.~Zhou, W.~Liu, {Fair resource
  allocation for system throughput maximization in mobile edge computing}, IEEE
  Access 6 (2018) 5332--5340.
\newblock \href {http://dx.doi.org/10.1109/ACCESS.2018.2790963}
  {\path{doi:10.1109/ACCESS.2018.2790963}}.

\bibitem{Pham2018a}
Q.-V.~V. Pham, T.~Leanh, N.~H. Tran, B.~J. Park, C.~S. Hong,
  \href{https://ieeexplore.ieee.org/document/8543561/}{{Decentralized
  Computation Offloading and Resource Allocation for Mobile-Edge Computing: A
  Matching Game Approach}}, IEEE Access 6 (2018) 75868--75885.
\newblock \href {http://dx.doi.org/10.1109/ACCESS.2018.2882800}
  {\path{doi:10.1109/ACCESS.2018.2882800}}.
\newline\urlprefix\url{https://ieeexplore.ieee.org/document/8543561/}

\bibitem{Haeringera}
G.~Haeringer, {Market design : auctions and matching}, The MIT Press, 2018.

\bibitem{Lyu2017a}
X.~Lyu, H.~Tian, C.~Sengul, P.~Zhang, {Multiuser joint task offloading and
  resource optimization in proximate clouds}, IEEE Trans. Veh. Technol. 66~(4)
  (2017) 3435--3447.
\newblock \href {http://dx.doi.org/10.1109/TVT.2016.2593486}
  {\path{doi:10.1109/TVT.2016.2593486}}.

\bibitem{Mao2016e}
Y.~Mao, J.~Zhang, K.~B. Letaief, {Dynamic Computation Offloading for
  Mobile-Edge Computing with Energy Harvesting Devices}, IEEE J. Sel. Areas
  Commun. 34~(12) (2016) 3590--3605.
\newblock \href {http://arxiv.org/abs/1605.05488} {\path{arXiv:1605.05488}},
  \href {http://dx.doi.org/10.1109/JSAC.2016.2611964}
  {\path{doi:10.1109/JSAC.2016.2611964}}.

\bibitem{Wang2016a}
Y.~Wang, M.~Sheng, X.~Wang, L.~Wang, J.~Li,
  \href{http://ieeexplore.ieee.org/document/7542156/}{{Mobile-Edge Computing:
  Partial Computation Offloading Using Dynamic Voltage Scaling}}, IEEE Trans.
  Commun. 64~(10) (2016) 4268--4282.
\newblock \href {http://dx.doi.org/10.1109/TCOMM.2016.2599530}
  {\path{doi:10.1109/TCOMM.2016.2599530}}.
\newline\urlprefix\url{http://ieeexplore.ieee.org/document/7542156/}

\bibitem{Tang2020}
Q.~Tang, L.~Chang, K.~Yang, K.~Wang, J.~Wang, P.~K. Sharma,
  \href{https://doi.org/10.1016/j.comcom.2019.12.018}{{Task number maximization
  offloading strategy seamlessly adapted to UAV scenario}}, Comput. Commun.
  151~(December 2019) (2020) 19--30.
\newblock \href {http://dx.doi.org/10.1016/j.comcom.2019.12.018}
  {\path{doi:10.1016/j.comcom.2019.12.018}}.
\newline\urlprefix\url{https://doi.org/10.1016/j.comcom.2019.12.018}

\bibitem{Han2017a}
Z.~Han, Y.~Gu, W.~Saad,
  \href{http://link.springer.com/10.1007/978-3-319-56252-0}{{Matching Theory
  for Wireless Networks}}, Springer International Publishing, 2017.
\newblock \href {http://dx.doi.org/10.1007/978-3-319-56252-0}
  {\path{doi:10.1007/978-3-319-56252-0}}.
\newline\urlprefix\url{http://link.springer.com/10.1007/978-3-319-56252-0}

\bibitem{Han2012}
Z.~Han, {Han Z., et al. Game Theory in Wireless and Communication Networks
  (CUP, 2011)(ISBN 9780521196963)(O)(553s){\_}GA{\_}.pdf} (2012).

\bibitem{Approach2003a}
A.~Abdulkadiroğlu, T.~S{\"{o}}nmez,
  \href{http://pubs.aeaweb.org/doi/10.1257/000282803322157061}{{School Choice:
  A Mechanism Design Approach}}, Am. Econ. Rev. 93~(3) (2003) 729--747.
\newblock \href {http://dx.doi.org/10.1257/000282803322157061}
  {\path{doi:10.1257/000282803322157061}}.
\newline\urlprefix\url{http://pubs.aeaweb.org/doi/10.1257/000282803322157061}

\bibitem{Ergin2003}
H.~Ergin, T.~S{\"{o}}nmez,
  \href{https://linkinghub.elsevier.com/retrieve/pii/S004727270500040X}{{Games
  of school choice under the Boston mechanism}}, J. Public Econ. 90~(1-2)
  (2006) 215--237.
\newblock \href {http://dx.doi.org/10.1016/j.jpubeco.2005.02.002}
  {\path{doi:10.1016/j.jpubeco.2005.02.002}}.
\newline\urlprefix\url{https://linkinghub.elsevier.com/retrieve/pii/S004727270500040X}

\bibitem{Sonmez2017}
C.~Sonmez, A.~Ozgovde, C.~Ersoy,
  \href{http://ieeexplore.ieee.org/document/7946405/}{{EdgeCloudSim: An
  environment for performance evaluation of Edge Computing systems}}, in: 2017
  Second Int. Conf. Fog Mob. Edge Comput., IEEE, 2017, pp. 39--44.
\newblock \href {http://dx.doi.org/10.1109/FMEC.2017.7946405}
  {\path{doi:10.1109/FMEC.2017.7946405}}.
\newline\urlprefix\url{http://ieeexplore.ieee.org/document/7946405/}

\bibitem{Cao2015}
Y.~Cao, P.~Hou, D.~Brown, J.~Wang, S.~Chen,
  \href{http://dl.acm.org/citation.cfm?doid=2757384.2757398}{{Distributed
  analytics and edge intelligence: Pervasive health monitoring at the era of
  fog computing}}, in: Proc. Int. Symp. Mob. Ad Hoc Netw. Comput., Vol.
  2015-June, ACM Press, New York, New York, USA, 2015, pp. 43--48.
\newblock \href {http://dx.doi.org/10.1145/2757384.2757398}
  {\path{doi:10.1145/2757384.2757398}}.
\newline\urlprefix\url{http://dl.acm.org/citation.cfm?doid=2757384.2757398}

\bibitem{Guo2017}
J.~Guo, B.~Song, Y.~He, F.~R. Yu, M.~Sookhak,
  \href{http://ieeexplore.ieee.org/document/7934315/}{{A Survey on Compressed
  Sensing in Vehicular Infotainment Systems}} (2017).
\newblock \href {http://dx.doi.org/10.1109/COMST.2017.2705027}
  {\path{doi:10.1109/COMST.2017.2705027}}.
\newline\urlprefix\url{http://ieeexplore.ieee.org/document/7934315/}

\end{thebibliography}

\end{document}